\documentclass[%
10pt,
notitlepage,
reprint,
a4paper,
superscriptaddress,
floatfix,
longbibliography
 amsmath,amssymb,
]{revtex4-2}

\usepackage{graphicx}
\usepackage{dcolumn}
\usepackage{amsmath}
\usepackage{amssymb}
\usepackage{graphicx}

\usepackage{bm}
\usepackage{ulem}
\usepackage[unicode=true,pdfusetitle,citecolor=blue,
urlcolor=blue,bookmarks=true,bookmarksnumbered=false,bookmarksopen=false,
 breaklinks=false,pdfborder={0 0 1},backref=false,colorlinks=true]
 {hyperref}
\usepackage{comment}
\usepackage{xcolor}
\usepackage{overpic}
\usepackage{xspace}
\usepackage{lmodern}

\usepackage{comment}
\usepackage{amssymb}
\usepackage{bm}
\usepackage{amsmath}
\usepackage{overpic}
\usepackage{xcolor}

\definecolor{myblue}{RGB}{12, 12, 158}
\definecolor{blue(ncs)}{rgb}{0.0, 0.53, 0.74}
\definecolor{crimson}{rgb}{0.86, 0.08, 0.24}
\definecolor{denim}{rgb}{0.08, 0.38, 0.74}
\definecolor{hanblue}{rgb}{0.27, 0.42, 0.81}
\definecolor{cadmiumorange}{rgb}{0.93, 0.53, 0.18}
\definecolor{darkblue}{rgb}{0,0,0.6}
\definecolor{darkred}{rgb}{0.6,0,0}

\definecolor{myred}{RGB}{158, 19, 22}
\definecolor{myorange}{RGB}{245, 150, 12}
\definecolor{mygreen}{RGB}{26, 118, 49}
\definecolor{Prune}{RGB}{99,0,60}
\definecolor{Purple}{RGB}{75, 0, 130}
\definecolor{Pink}{RGB}{255, 105, 180}
\definecolor{deepskyblue}{RGB}{0, 191,255}
\definecolor{limegreen}{RGB}{50, 205, 50}
\definecolor{crimson}{rgb}{0.86, 0.08, 0.24}
\definecolor{coral}{rgb}{1.0, 0.5, 0.31}
\newcommand{\caja}[1]{\left[ {#1} \right] }
\newcommand{\paren}[1]{\left( {#1} \right) }
\newcommand{\lazo}[1]{\left\{ {#1} \right\} }
\newcommand{\av}[1]{\left\langle {#1} \right\rangle }

\newcommand{\figpanel}[1]{(\textbf{\lowercase{#1}})}

\definecolor{blue(ncs)}{rgb}{0.0, 0.53, 0.74}

\newcommand{\panel}[1]{(\textbf{\lowercase{#1}})}

\newcommand{\Nv} {{N_\mathrm{v}}}
\newcommand{\Nh} {{N_\mathrm{h}}}

\usepackage{amsmath}
\usepackage{comment}

\begin{document}

\preprint{APS/123-QED}

\title{Inferring effective interactions and task-related brain states from large-scale neural activity with Restricted Boltzmann Machines}

\author{Nicolas Béreux}
\affiliation{Université Paris-Saclay, CNRS, INRIA, LISN, Gif-sur-Yvette, France}
\author{Giovanni Catania}
\email{gcatania@ucm.es}
\affiliation{Departamento de Física Teórica, Universidad Complutense de Madrid, 28040
Madrid, Spain.}
\affiliation{Institute for Cross-disciplinary Physics and Complex Systems IFISC (CSIC-UIB),
Campus Universitat Illes Balears, 07122 Palma de Mallorca, Spain.}

\author{Aurélien Decelle}
\affiliation{Escuela Técnica Superior de Ingenieros Industriales, Universidad Politécnica de Madrid, 28006 Madrid, Spain.
}
\affiliation{GISC - Grupo Interdisciplinar de Sistemas Complejos 28040 Madrid, Spain.}

\author{Francesca Mignacco}
\affiliation{Joseph Henry Laboratories of Physics and Lewis–Sigler Institute, Princeton University, Princeton NJ 08544 USA}
\affiliation{Initiative for the Theoretical Sciences, The Graduate Center, City University of New York, 365 Fifth Ave, New York NY 10016 USA}
\author{Alfonso de Jesús Navas Gómez}
\affiliation{Departamento de Física Teórica, Universidad Complutense de Madrid, 28040
Madrid, Spain.}

\author{Beatriz Seoane}
\email{beseoane@ucm.es}
\affiliation{Departamento de Física Teórica  \& IPARCOS, Universidad Complutense de Madrid, 28040 Madrid, Spain.}
\affiliation{GISC - Grupo Interdisciplinar de Sistemas Complejos 28040 Madrid, Spain.}

\begin{abstract}Large-scale electrophysiological recordings now enable the simultaneous monitoring of thousands of neurons across multiple brain regions, revealing structured variability in population activity. Understanding how such collective patterns emerge from microscopic interactions requires models that are both scalable and interpretable. Here, we use Restricted Boltzmann Machines (RBMs) to model the activity of $\sim1500$--$2000$ simultaneously recorded neurons from the Allen Institute for Brain Science Visual Behavior Neuropixels dataset. Compared with conventional pairwise maximum-likelihood approaches, RBMs require substantially fewer parameters, can be trained within minutes, and accurately reproduce higher-order and global empirical statistics. They remain interpretable by defining effective interactions between neurons, including higher-order terms, and uncover dominant coordination patterns that reflect anatomical organization. Analysis of the learned free-energy landscape further identifies task-related brain states, while Monte Carlo sampling captures the global relaxation dynamics observed in the data. These results establish RBMs as efficient, scalable, and interpretable tools for extracting statistical structure from large-scale neural recordings and linking collective neural activity to brain organization and behavior.
\end{abstract}

\date{\today}
\maketitle

\section{Introduction}

Recent advances in large-scale electrophysiology, particularly Neuropixels probes, now enable simultaneous recordings from thousands of neurons across multiple brain regions \cite{doi:10.1126/science.abf4588}, revealing high-dimensional and structured population dynamics linked to perception, internal states, and behavior. A central challenge is to develop statistical frameworks capable of capturing and interpreting such collective activity.
Within equilibrium statistical mechanics, neural activity can be modeled as samples from an unknown high-dimensional distribution encoding effective neuronal interactions. Maximum-entropy methods~\cite{nguyen_inverse_2017,meshulam2025statistical}, typically based on firing rates and pairwise correlations, reduce this problem to inverse Ising inference and have proven successful for small populations ($N\sim100$). However, their quadratic parameter scaling makes them impractical for modern large-scale recordings. Recent approaches mitigate this issue by constraining the interaction topology (e.g. on trees ~\cite{lynn2023exact}, improving scalability at the cost of excluding potentially relevant structures such as feedback loops. Some extentions have been propoosed to account for the presence of short loops in planar structures~\cite{carcamo2025statistical} but the topology selection comes at an additional computational cost and limits the resulting inference to sparse pairwise correlations arranged on a fixed topology.

Beyond parameter scaling, restricting models to pairwise interactions is a major limitation, as it neglects higher-order structure in population activity; in maximum-entropy approaches, this corresponds to encoding only two-body correlations. Higher-order interactions have been shown to reproduce empirical statistics \cite{ohiorhenuan2010sparse,Ganmor2011pnas,chelaru2021high,di2025extended}, capture collective phenomena such as synchronization \cite{tkavcik2014searching,shahidi2019high}, and enhance the information storage capacity of artificial neural networks \cite{NIPS2016_eaae339c,ramsauer_hopfieldallyouneed,Krotov2023review,lucibello_mezard_exphopfield}. More fundamentally, enforcing only firing rates and pairwise correlations may not provide the most compact or expressive description of neural activity. Identifying principled alternatives that capture higher-order statistics without uncontrolled parameter growth remains an open problem. 
In parallel, modern generative models can learn complex high-dimensional structure directly from data and generate realistic samples across domains ranging from vision to protein structure. However, they typically require large datasets and remain difficult to interpret. 

RBMs provide a natural compromise between maximum-entropy models and deep architectures: they retain analytical tractability~\cite{decelle2021exact}, are universal approximators \cite{leroux_power_of_RBMs}, and offer a parsimonious representation with far fewer parameters than pairwise models while still capturing higher-order statistics through latent variables. Their architecture also enables efficient parallel sampling, making them well suited for large neural populations. Consistently, RBMs have been shown to reproduce  mean activity and pairwise correlations in recordings of up to tens of thousands of neurons \cite{van2023neural,cossioLaRBM}, while revealing functional organization and supporting decoding and dynamical modeling \cite{yang26cross,monnens24the}.

\begin{figure*}[t!]
    \centering
\includegraphics[trim=2.5cm 5.5cm 2cm 3cm, width=0.9\textwidth]{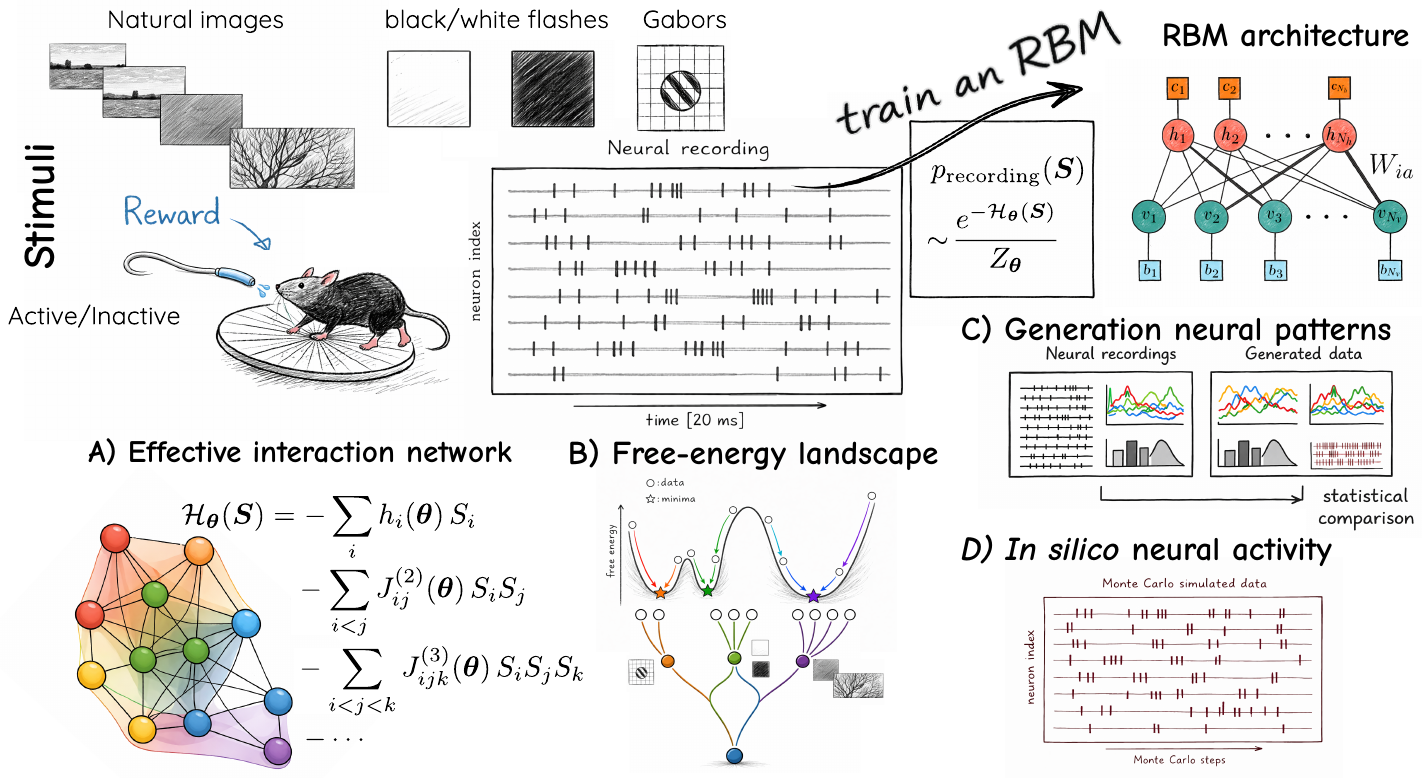}
    \caption{\textbf{Pipeline of the analysis.} We analyze datasets coming from the Visual Behavior Neuropixels project, comprising large-scale neural recordings from mice across three experimental epochs: active behavior (visual change-detection task), receptive field characterization (Gabor stimuli and full-field flashes), and passive replay of the same stimulus sequence. Each recording is annotated by stimulus identity and behavioral condition (active vs passive).
We train RBMs to reproduce the empirical statistics of population activity, treating each time point as an independent sample from an underlying distribution. The learned models are then used in several complementary ways: (A) mapping the energy function onto an effective spin Hamiltonian with interactions of arbitrary order, providing access to the inferred functional connectivity; (B) analyzing the minima of the free-energy landscape to cluster neural snapshots into interpretable states associated with stimuli or behavioral context, despite the absence of labels during training; and generating neural patterns via Markov Chain Monte Carlo (MCMC) sampling, either to compare empirical and model statistics (C) or to produce \textit{in silico} neural dynamics for direct comparison with the observed temporal activity (D).
    }
    \label{fig:pipeline}
\end{figure*}
In recent years, the architectural simplicity of RBMs has made them effective and interpretable inference engines, particularly for neuroscience applications. Hidden units can be linked to biological function in protein families~\cite{tubiana_couplings_elife} or trained to select specific features~\cite{fernandez2023disentangling} or structural conformations~\cite{cossio_switches}. Moreover, RBMs can be mapped onto multibody Ising models (Fig.~\ref{fig:pipeline}A), enabling the direct inference of effective interactions, including higher-order couplings, from data~\cite{cossu2019machine,decelle2024inferring,decelle2025inferringhighordercouplingsneural}. Their tractability also permits the use of tools from the statistical physics of disordered systems, such as the TAP equations~\cite{maillard2019high,tramel2018deterministic,decelle2023unsupervised}, to characterize the free-energy landscape.  This provides a principled way to identify relevant patterns and cluster samples according to their proximity in the learned representation (Fig.~\ref{fig:pipeline}B), uncovering hidden structure and enabling unsupervised classification~\cite{decelle2023unsupervised}. Finally, recent advances in maximum-likelihood training and optimized MCMC sampling~\cite{bereux2024fasttrainingsamplingrestricted,bereux2026ptt} have made RBMs scalable to large datasets while preserving an accurate description of complex multimodal distributions and enabling reliable likelihood estimation.

In this work, we apply RBMs to describe the statistics of the Neuropixels recordings from thousands of neurons across cortical and subcortical regions in mice performing visually guided task~\cite{AllenData_behavior,bennett2026map}. 
We show that RBMs are able to reproduce accurately the overall statistics of real recordings, including high order and non-linear ones. We further employ RBMs as a tool to infer neural interaction structure in a manner analogous to maximum entropy models, but without predefined constraints or manual topology selection. We further exploit the free-energy landscape to extract task-related patterns by linking its minima to external stimuli and behavioral states, and show that Monte Carlo dynamics on this methods are able to catch the overall relaxations in the system. Figure~\ref{fig:pipeline} summarizes the pipeline of the present work.

\section{Restricted Boltzmann Machines and their mapping to multibody interacting models}
Restricted Boltzmann machines (RBMs)~\cite{Smolensky} are probabilistic energy-based generative models naturally framed within equilibrium statistical mechanics. They consist of a bipartite architecture with \textit{visible} variables (data) and \textit{hidden} variables (latent degrees of freedom), coupled only between layers. The model defines a Boltzmann--Gibbs distribution over joint configurations, where hidden units encode dependencies among visible variables and, upon marginalization, induce effective pairwise and higher-order correlations.
Here we consider RBMs with binary visible and hidden units, $v_i,h_a \in \{0,1\}$. For visible variables, this choice provides a minimal representation of neuronal activity as a two-state process (active/silent) within a given time window. For hidden variables, binary parametrization ensures universal approximation capabilities~\cite{leroux_power_of_RBMs}. The Hamiltonian of the model is then given by
\begin{equation}
H_{\bm{\theta}}(\bm{v},\bm{h}) =
- \sum_{i,a} W_{ia} v_i h_a
- \sum_i b_i v_i
- \sum_a c_a h_a ,
\label{eq:RBMHamiltonian}
\end{equation}
where $b_i$ and $c_a$ denote local fields (biases) acting on visible and hidden units, respectively, and $W_{ia}$ are the interaction strengths. The parameter dependence is summarized by $\bm{\theta} \equiv (\bm{W},\bm{b},\bm{c})$. A schematic representation of this architecture is shown in Fig.~\ref{fig:pipeline} (top-right). The corresponding Boltzmann distribution reads
\begin{equation}
p_{\bm{\theta}}(\bm{v},\bm{h}) =
\frac{1}{Z_{\bm{\theta}}}
e^{-H_{\bm{\theta}}(\bm{v},\bm{h})},
\qquad
Z_{\bm{\theta}} =
\sum_{\bm{v},\bm{h}}
e^{-H_{\bm{\theta}}(\bm{v},\bm{h})},
\label{eq:BoltzmannDistributionRBM}
\end{equation}
where $Z_{\bm{\theta}}$ is the partition function. 
The model is trained (see the Materials and Methods section) so that its marginal distribution over visible variables,
\begin{equation}
p_{\bm{\theta}}(\bm{v})=
\frac{1}{Z_{\bm{\theta}}}
\sum_{\bm{h}}
e^{-H_{\bm{\theta}}(\bm{v},\bm{h})},
\label{eq:pmarginal}
\end{equation}
approximates the empirical data distribution,
$p_{\mathrm{emp}}(\bm{v})\!=\!\frac{1}{M}\sum_{\mu=1}^M \delta\paren{\bm{v}^{(\mu)}\!-\!\bm{v}}$, where $\{\bm{v}^{(\mu)}\}_{\mu=1}^M$ are the $M$ training samples. The number of hidden units is chosen as the smallest value that maximizes the log-likelihood of a held-out \textit{test set}, which is not used during training. Increasing the number of hidden units beyond this point does not improve the test log-likelihood and instead leads to pronounced overfitting (see Supporting Information, SI). We find that $N_\mathrm{h}=256$ is an overall optimal value for different experimental sessions, each one corresponding to a different animal subject (additional details are given in Section 1 of the SI~\cite{Supplementary}).  
All RBMs considered in the following are trained using the algorithm introduced in~\cite{bereux2026ptt} (see the Materials and Methods section), with an average training time of $30$--$45$ minutes on a single NVIDIA RTX 3090 GPU (additional runtime details are given in Section II of the SI). However, the maximum likelihood point is typically reached at about $1/10$ of the full training time. 
\subsection*{Mapping}
Marginalizing over the hidden units yields an explicit energy-based model for the data distribution, enabling RBMs to capture complex probability landscapes. The marginal over visible variables \eqref{eq:pmarginal} can be written as a Boltzmann distribution, $p(\bm v)=e^{-\mathcal{H}_{\bm{\theta}}(\bm v)}/Z_{\bm{\theta}}$, where $\mathcal{H}_{\bm{\theta}}(\bm v)$ is the effective energy function over visible configurations:
\begin{equation}
    \mathcal{H}_{\bm{\theta}}\paren{\bm v} = - \sum_{i} b_i v_i - \sum_a \ln \left( 1 +  e^{c_a + \sum_i W_{ia} v_i }\right). \label{eq:energy_visible}
\end{equation}
As shown in Refs.~\cite{decelle2024inferring,decelle2025inferringhighordercouplingsneural}, this Hamiltonian can be rewritten exactly as a generalized Ising model with interactions of arbitrary order
\begin{align}
\!\!\!\textstyle\mathcal{H}_{\bm{\theta}}(\boldsymbol{s}) &\!=\! -\! \sum_i H_i(\bm{\theta}) s_i\! - \!\sum_{i < j}\!\! J_{ij}^{(2)}(\bm{\theta}) s_i s_j\nonumber\\ &\!\!\!-\!\! \!\!\sum_{i < j<k}\!\!\! J_{ijk}^{(3)}(\bm{\theta}) s_i s_j s_k\!-\!\! \!\!\sum_{i < j<k<l}\!\!\! J_{ijkl}^{(4)}(\bm{\theta}) s_i s_j s_k s_l
\! +\! \dots,\!\!\! \label{eq:Generalized_ising_model}
\end{align}
where $s_i \in \{\pm 1\}$ are Ising spin variables, defined as $s_i \equiv 2v_i-1$. This Ising expansion is equivalent to a discrete Fourier (Walsh--Hadamard) expansion and therefore provides an irreducible representation, corresponding to the orthogonal basis on the Boolean hypercube, unlike an expansion in the original $\{0,1\}$ variables. As a result, it concentrates most of the interaction information into the lowest-order terms~\cite{decelle2026distributional}, leading to substantially more accurate interaction inference~\cite{decelle2024inferring,decelle2025inferringhighordercouplingsneural}.
This representation enables the inference of interaction networks beyond the pairwise structure captured by conventional maximum entropy models. Higher-order interactions are inferred implicitly using only $O(N_\mathrm{v}N_\mathrm{h})$ parameters, avoiding the combinatorial increase associated with modeling them explicitly. We can also quantify the aggregate contribution of higher-order interactions involving each pair of neurons using the $R_{ij}^{(2)}$ measure. Further details on it, the extraction of effective couplings and the training procedure are provided in the Materials and Methods section.
\begin{figure*}[t!]
    \centering
    \begin{overpic}[width=\textwidth]{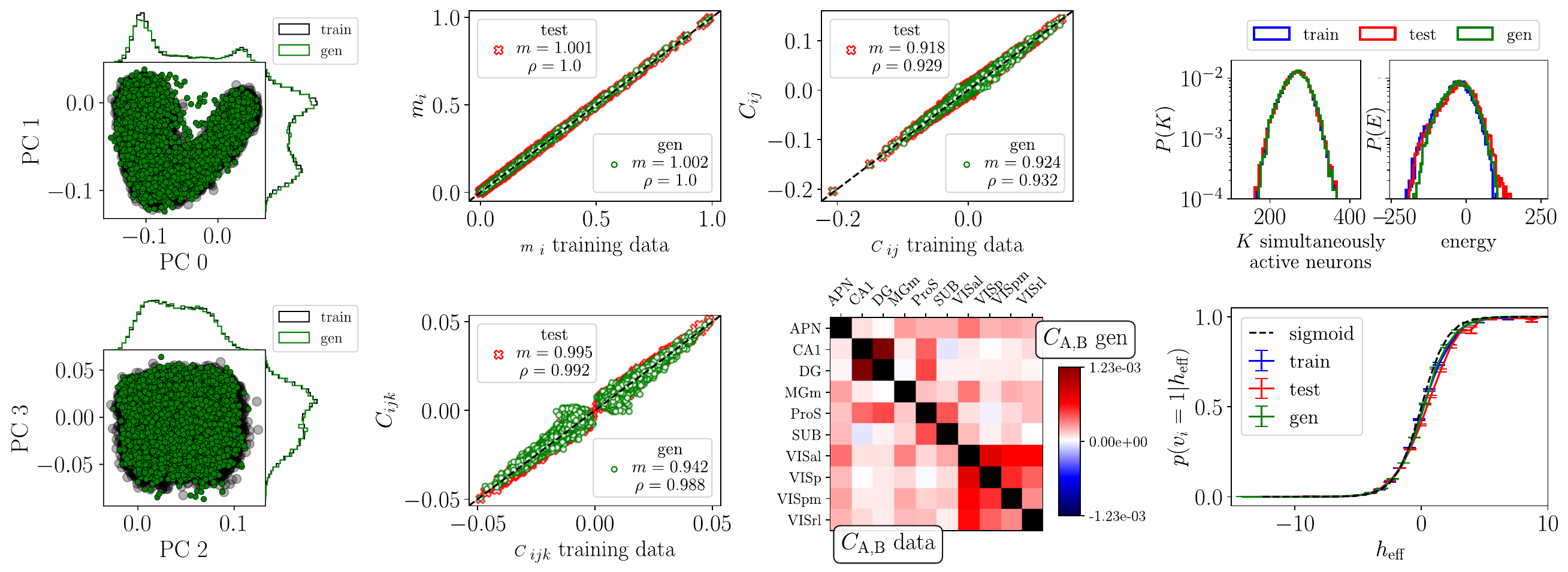}
    \put(1,33){{\figpanel{a}}}
    \put(1,15){{\figpanel{b}}}
    \put(24,35){{\figpanel{c}}}
    \put(47,35){{\figpanel{d}}}
    \put(24,15){{\figpanel{e}}}
    \put(47.5,15){{\figpanel{f}}}
    \put(78.8,30.5){{\figpanel{g}}}
    \put(95,30.5){{\figpanel{h}}}
    \put(75,18){{\figpanel{i}}}
\end{overpic}
    \caption{ \textbf{Generated versus real data statistics.} Quality of generated samples assessed by comparing empirical statistics of $N=2028$ neurons with those of the training and test sets, using $10{,}000$ randomly selected sequences from each of the training, test, and generated datasets. Panel \figpanel{a}: projection of training data (in black) and generated samples (in green) along the first two principal components (PCA directions) of the training set of data;  histograms at the top and at the right highlight the one-dimensional marginal distributions over the first and second direction, respectively. \figpanel{b}: same projection but on the 2-dimensional space spanned by the third and fourth PCA directions of the data. Panel \figpanel{c}: single-neuron firing rate; \figpanel{d}: neuron-neuron covariance matrix; \figpanel{e}: three-body correlations, displayed on the union set of largest $10^5$ correlations (in absolute value) of the three sets;  Red points denote training–test comparisons; green points denote training–generated comparisons. The slope of a linear fit and the Pearson correlation coefficient are reported in the legend. Panel \figpanel{f}: coarse grained connectivity between brain areas, comparison between training data (left-lower triangular part) and generated data (right-upper triangular part).
Panel \figpanel{g}: histogram of the number of simultaneously active neurons; \figpanel{h}: histogram of sample energies assigned by the RBM; \figpanel{i}: probability that a neuron is active conditioned on the network state: points are obtained by averaging over bins with similar values of $h_\mathrm{eff}$ (defined in \eqref{eq:heff}); error bars correspond to the standard error of the mean.
    }
    \label{fig:genstats}
\end{figure*}

\section{Results}

\subsection{Reproducing high order statistics in the data}
First, we analyze the ability of the trained models to accurately reproduce different data statistics. 
We consider trainings performed on a total of 8 experimental sessions, each one corresponding to a different mouse. Trainings are performed with the same conditions and hyper-parameters (see Material and Methods and Section II of the SI). The results of this section shown in Fig.~\ref{fig:genstats} refer to one of them (mouse ID 560771). We report the statistics for one additional session in the Appendix; the remaining sessions show essentially the same quantitative agreement between generated and training data and are therefore not shown.

The first two panels \figpanel{a}-\figpanel{b} show the projections of both the training data and the generated samples onto the first principal components of the data, with two different $2d$ projections. The two empirical distributions closely match, including in the one-dimensional projections shown as the top and right histograms in \figpanel{a}-\figpanel{b}. The same behavior is observed along the other PCA directions (not shown). The panels \figpanel{c}-\figpanel{d}-\figpanel{e} show the scatter plots of first, second and third centered moments ---that is, single neuron average firing rates, 2-point and 3-point connected correlations--- respectively: these quantities are computed for training, test and generated samples and scatter-plotted against the first. 
We can see a perfect agreement between these sets of statistics, as further quantified by the Pearson correlation coefficient $\rho$ and the slope of the linear curve fit $m$ between each set of statistics (values are shown inside each panel): similar values of both $m,\rho$ are found for the pairs train/test data and train/generated data, confirming that the RBM is able to generate samples statistically indistinguishable from the ones in the test-set.
Such comparisons are performed using the same number $M=10^4$ samples for each set. 
The agreement between data and generated statistics is also evident at a more coarse-grained level, as clear from \figpanel{f}, where we plot the static correlation between pairs of brain areas (see SI) for the $10$ areas with the most recorded neurons within, playing a similar role to functional connectivity estimates~\cite{Stevenson2008-el}. 
Another example of generation quality is shown in \figpanel{g}–\figpanel{h}, where we report the histograms of the number of simultaneously active neurons, $P\paren{K}$ (in \figpanel{g}), and the histogram of the RBM energies, $P\paren{E}$ (in \figpanel{h}, computed on the visible nodes, i.e., using $\mathcal{H}_{\bm \theta}(\bm v)$ from \eqref{eq:energy_visible}) assigned to each configuration in the three sets of samples (training, test, and generated).
 The comparison in \figpanel{h} measures the ability of the model to assign low-energy configurations at most likely patterns of neural activity, and conversely high energy to poorly observed ones. Both curves perfectly agree in the central part of the histograms, with difference in tail due to finite-sampling fluctuations effect. 
Finally, we show in \figpanel{i} the predicted conditional probability of a single neuron being active conditioned to the rest of the network's state. This measure has been first considered in ~\cite{tkavcik2014searching} and it is a signal of the ability of the model to capture the influence that each neuron is subject to by the rest of the network. The effective fields for each neuron $i$ and each sample $\mu$ are computed from the energy difference between two configurations in which only neuron $i$ changes state, that is
\begin{align}
    h_i^{\mathrm{eff}} = & \, \mathcal{H}_\theta \paren{v_1, \ldots, v_{i-1}, v_i=1, v_{i+1}, \ldots, v_\Nv} \nonumber \\ 
    & - \mathcal{H}_\theta \paren{v_1, \ldots, v_{i-1}, v_i=0, v_{i+1}, \ldots, v_\Nv}  \label{eq:heff}
\end{align}
where $\mathcal{H}$ is the visible nodes' energy computed from the trained EBM through 
    \eqref{eq:energy_visible}. This quantity is computed, binned according to $h_i^{\mathrm{eff}}$, and plotted for the three sets of samples (train, test, and generated). The results are then compared with a parameter-free sigmoid fit:
    \begin{equation}
        p\paren{v_i=1 \mid h_i^{\mathrm{eff}}} = \frac{1}{1 + e^{- h_i^{\mathrm{eff}} }}\,.
    \end{equation}
It is straightforward to verify that this relation is trivially satisfied for data distributed according to $p_{\bm \theta}(\bm v)$; it therefore provides a measure of how well the training and test data are described by the model distribution. The three binned curves perfectly match with each other and overlap with the sigmoid trend in Fig.~\ref{fig:genstats}-\figpanel{i}. 

Finally, these results are consistent across RBMs trained on different experimental sessions (i.e., different animal subjects), see the SI~\cite{Supplementary}.

\subsection{Analysis of effective model}
\begin{figure*}[t!]
    \centering
    \begin{overpic}[width=\textwidth]{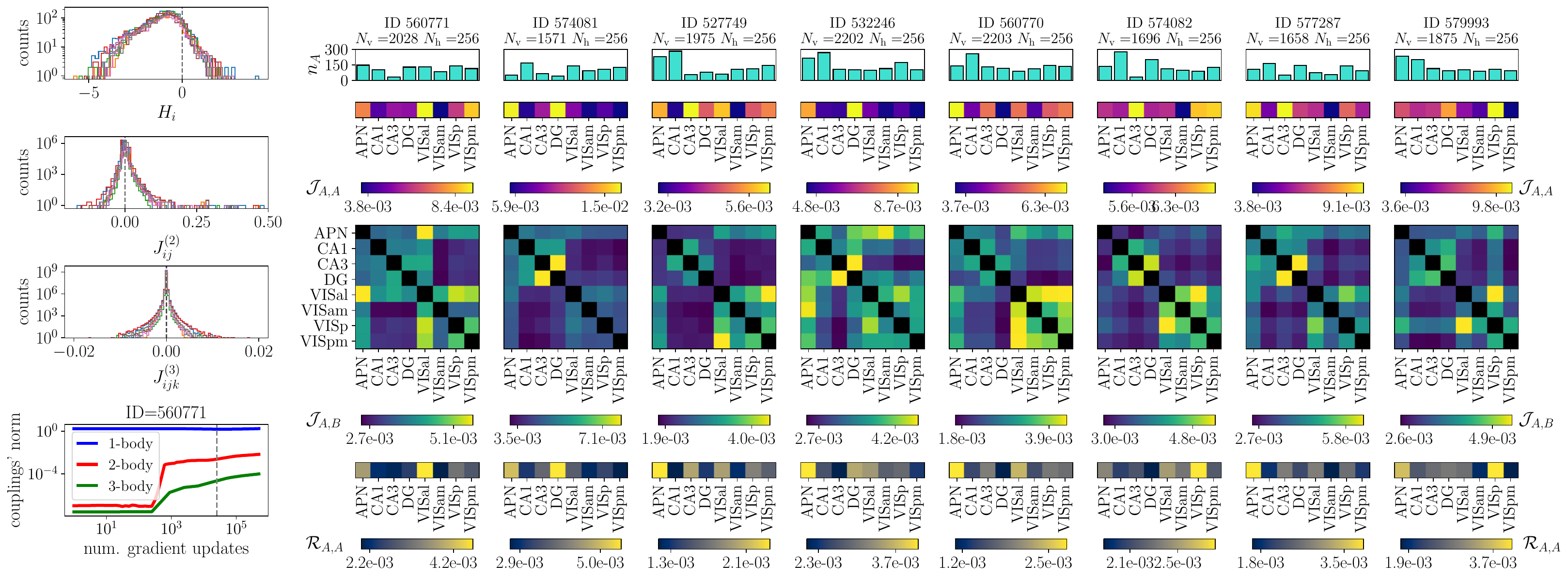}
    \put(4.3,35){{\figpanel{a}}}
    \put(4.3,26.5){{\figpanel{b}}}
    \put(4.3,18.5){{\figpanel{c}}}
    \put(4,11){{\figpanel{d}}}
    \put(13,11){{$*$}}    
    \put(28.8,36){{$*$}}    
    \put(97,35){{\figpanel{e}}}
    \put(97,29){{\figpanel{f}}}
    \put(97,18.5){{\figpanel{g}}}
    \put(97,6.5){{\figpanel{h}}}
\end{overpic}
    \caption{ \textbf{Extraction of effective model}. \figpanel{a}-\figpanel{b}-\figpanel{c}: histograms of effective fields $H_i$ (in \figpanel{a}), two-body interactions $J_{ij}^{(2)}$ (in \figpanel{b}) and three-body interactions $J_{ijk}^{(3)}$ (in \figpanel{c}) computed from the trained RBMs on each dataset for the $8$ experimental sessions / mice. \figpanel{d}: evolution of the (normalized, see SI) Frobenius norm of each of the $3$ coupling set vs training time, for one of the training (ID 560771, corresponding to the first column in the following panels); a vertical line highlights the maximum (test) likelihood early-stopping point. \figpanel{e}: number of neurons sampled in each brain region common to the $8$ mice considered. \figpanel{f}: heatmap of the coarse-grained diagonal interaction $\mathcal{J}_{A,A}$ within a brain area $A$. \figpanel{g}: heatmap of the coarse-grained off-diagonal interaction $\mathcal{J}_{A,B}\, A \neq B$. \figpanel{h}: heatmap of the coarse-grained diagonal coefficients $\mathcal{R}_{A,A}$ quantifying the role of high-order interactions between neurons of area $A$. In all the panels, the effective models are evaluated at the training time maximizing the test log-likelihood. \label{fig:effective_model} 
    }
\end{figure*}

Once the model is trained, we can use Eq.~\ref{eq:energy_visible} (see also Materials and Methods section) can be used to extract an effective microscopic model of the recorded neuronal population and, in principle, infer interactions among arbitrary subsets of neurons. For simplicity, we restrict our analysis explicitly to higher-order interactions $J_{i_1\ldots i_n}^{(n)}$ up to third order, or to their aggregate contribution beyond second order, quantified by the $R_{ij}^{(2)}$ measure. 
The following results, summarized in Fig.~\ref{fig:effective_model}, compare models trained on eight experimental sessions, each corresponding to a different mouse. Trainings are performed with the same conditions and hyperparameters (see Material and Methods and Section II of the SI). Panels \figpanel{a}-\figpanel{b}-\figpanel{c} display the histograms of effective fields, $2-$body and $3-$body interactions of the trained RBM, computed at the training point that maximizes the test log-likelihood. Both $2-$body and $3$-body couplings are strongly peaked around $0$ (note that the vertical axis is in log-scale), signaling an effective sparse network with a tail of stronger interactions, mirroring the sparse heavy-tailed organization found across species in both synaptic connectivity and activity covariances~\cite{lynn2024heavy}. Furthermore, the distribution of $2-$body couplings in \figpanel{b} is partially skewed towards positive values. In each of these panels, each curve corresponds to different experiments: all curves are overlapping significantly, confirming that the RBM captures similar distributions of microscopic interactions across different mouses.
In \figpanel{d}, we further show the evolution of the norms of the three sets of interactions during training, computed as normalized Frobenius norms (see SI), for a single training run (mouse ID=560771, also used in Fig.~\ref{fig:genstats}). Their stabilization at late stages of training qualitatively indicates that the overall strength of the inferred interactions becomes stable over the course of training.

\begin{figure}[t!]
    \centering
\begin{overpic}[width=\columnwidth]{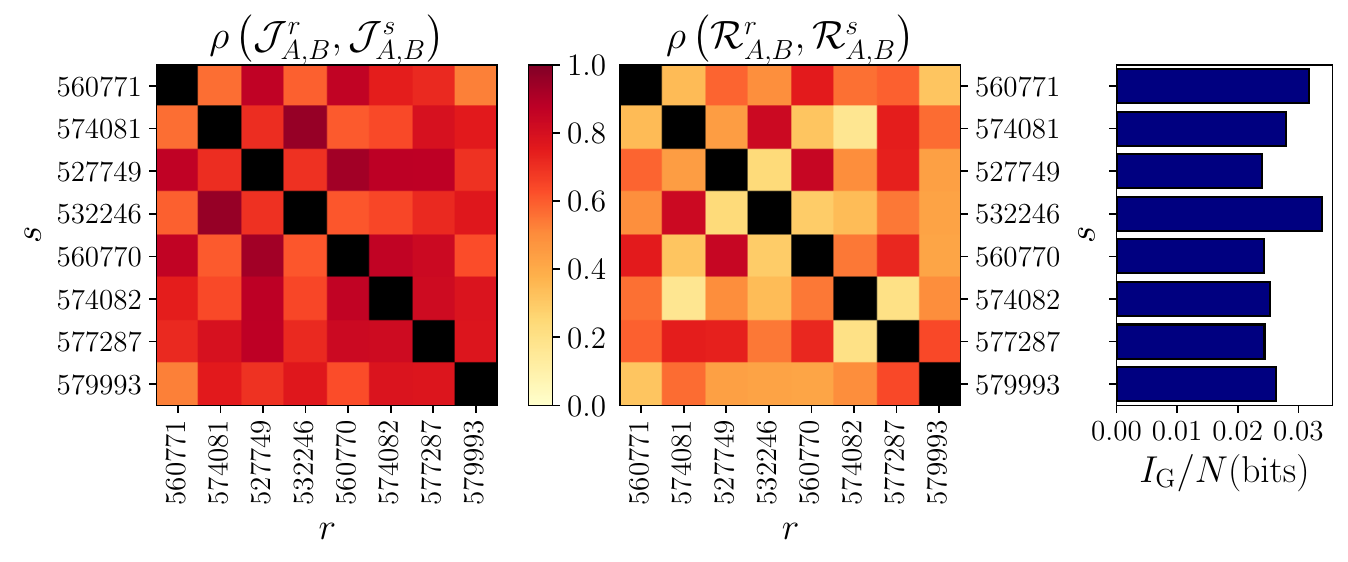}
    \put(8,39){{\figpanel{a}}}
    \put(43,39){{\figpanel{b}}}
    \put(85,39){{\figpanel{c}}}
\end{overpic}
    \caption{Pearson correlation coefficient between the set of coarse-grained interactions $\mathcal{J}^r_{A,B}$ (for $A\geq B$) in \figpanel{a} and of the coarse-grained coefficients $\mathcal{R}^r_{A,B}$ in \figpanel{B}, quantifying the aggregated high-order inferred correlations between pairs of brain areas. The coefficient map is computed between all pairs of mice $r\neq s$. Panel \figpanel{c}: information per neuron gained with the RBM w.r.t. an independent model fitting only the first moments of the data, measured in bits (per neuron) for each experimental session.\label{fig:comparison_rats}}
\end{figure}

The recorded metadata for each experimental session further provide information about which brain area in the mouse's cortex each recorded neuron belongs to. Therefore, we can compute a coarse-grained set effective 2-body interactions, both \textit{within} a single brain area and \textit{between} pairs of areas:
\begin{equation}
    \mathcal{J}_{A,B} = \sqrt{ \frac{1}{n_A \paren{n_B - \delta_{A,B}}}\sum_{i \in A, j\in B} \paren{J^{(2)}_{ij}}^2 } \label{eq:Jij_ba_formula}
\end{equation} where $n_A = \sum_{i=1}^\Nv \mathbb{I}\caja{i \in A}$ counts, the number of elements in the area $A$. The above formula allows to compute which areas are collectively more interacting between each other during the experiment. This measure differs from standard functional connectivity estimates, as it is not directly based on cross-correlations between brain areas but rather on direct coupling analysis estimated through the trained RBM. Furthermore, this measure enables direct comparison across mice, unlike single-neuron effective interactions, since individual neurons cannot be mapped one-to-one across animal subjects.
The result is displayed in Fig.~\ref{fig:effective_model}-\panel{f}-\panel{g}, where we show the coarse-grained interaction map for all the $8$ mice (one per column), on a restricted set of brain areas whose the number of recorded neurons is $\geq 30$ for each mice. The coarse-grained interaction map is shown separately for the diagonal components $\mathcal{J_{A,A}}$ in \figpanel{f}, and for the off-diagonal (symmetric) components $\mathcal{J}_{A,B}$ in \figpanel{g}.  It is qualitatively clear that for all the trainings, stronger effective interactions are captured between neurons belonging to visual cortex areas, consistent with the type of (visual) task carried out in the experimental setup from which data are collected. Furthermore, these results highlight modularity in both the visual cortex and the hippocampal areas (CA1, CA3, DG): this is consistent both with the type of task (guided by a visual stimulus) and with the mouse being familiar with the experimental setup, thus involving memory circuits within the hippocampus. The normalization factor in \eqref{eq:Jij_ba_formula} further allows for a more fair comparison because the same region is sampled in different ways across different subjects, as shown in panels \figpanel{e} which report the number of recorded neurons into each of the common brain areas for each experiment.
We report in the Appendix further results about the distributions of inferred interaction strengths $J_{ij}$ for all the pairs of neurons inside a given region, and comparisons are made across multiple regions, revealing a significant heterogeneity in the distribution of effective interactions, with some regions clearly displaying a more fat-tailed distribution of couplings than others. 
Finally, we assess the importance of higher-order interactions in Fig.~\ref{fig:effective_model}-\figpanel{h}, where we show the mean $R_{ij}^{(2)}$ within each brain region, coarse-grained as in the previous panels. This quantity, defined in Eq.~(\ref{eq:R2_def}) (see the Materials and Methods section), quantifies the aggregate strength of higher-order interactions beyond second order involving neurons $i$ and $j$.
The quantity shown is simply the mean across the coefficients within each area, namely $\mathcal{R}_{A,A} = \binom{n_A}{2}^{-1} \sum_{i<j; i,j \in A} R_{ij}^2 $.  We can observe that some regions exhibit significantly larger values of $R_{ij}^{(2)}$ than others, highlighting the stronger role of higher-order interactions in those areas. However, it remains unclear whether this reflects an intrinsic property of the region or simply the density of neurons sampled there, as suggested in previous works~\cite{meshulam2021successes,meshulam2025statistical}. In particular, regions showing stronger higher-order interactions are not very consistent across different mice. 
To quantitatively compare the inferred interactions across mice, we compute the Pearson correlation coefficient between the vectors of diagonal and off-diagonal coarse-grained couplings for each pair of mice $r$ and $s$. The result is shown in Fig.~\ref{fig:comparison_rats}, both for the two-body interactions $\{ \mathcal{J}_{A,B}, A \geq B \}$ in \figpanel{a} and for the matrix of R coefficients $\{ \mathcal{R}_{A,B}, A \geq B \}$ in \figpanel{b}. Overall, we observe clear correlations between the effective interactions inferred across different pairs of experiments, with an average Pearson correlation of $0.74$, suggesting that well-trained RBMs capture relevant microscopic information consistently across animal subjects.
Weaker correlations can instead be noted for the set of high-order interactions, suggesting that these might be dominated by sampling noise or simply being much weaker to be reliably estimated from these datasets.
Finally, in Fig.~\ref{fig:comparison_rats}-\figpanel{c}, we show the information gained by the trained RBM relative to an independent model (equivalently, an RBM with null weights) that reproduces only the first moments of the data, and corresponds to the initialization of our models. This measure (defined in Sec.~\ref{sec:MMinformationgain}) was previously introduced in~\cite{carcamo2025statistical} to quantify the information gained by inferring a model that captures correlations in the data. Across the experimental sessions considered here, we find comparable, and in some cases larger, values than those reported in~\cite{carcamo2025statistical}, reaching more than $0.03$ bits per neuron.

\subsection{Unsupervised classification of neural patterns using the free-energy landscape}
\begin{figure*}[t!]
    \centering
\begin{overpic}[width=0.9\textwidth]{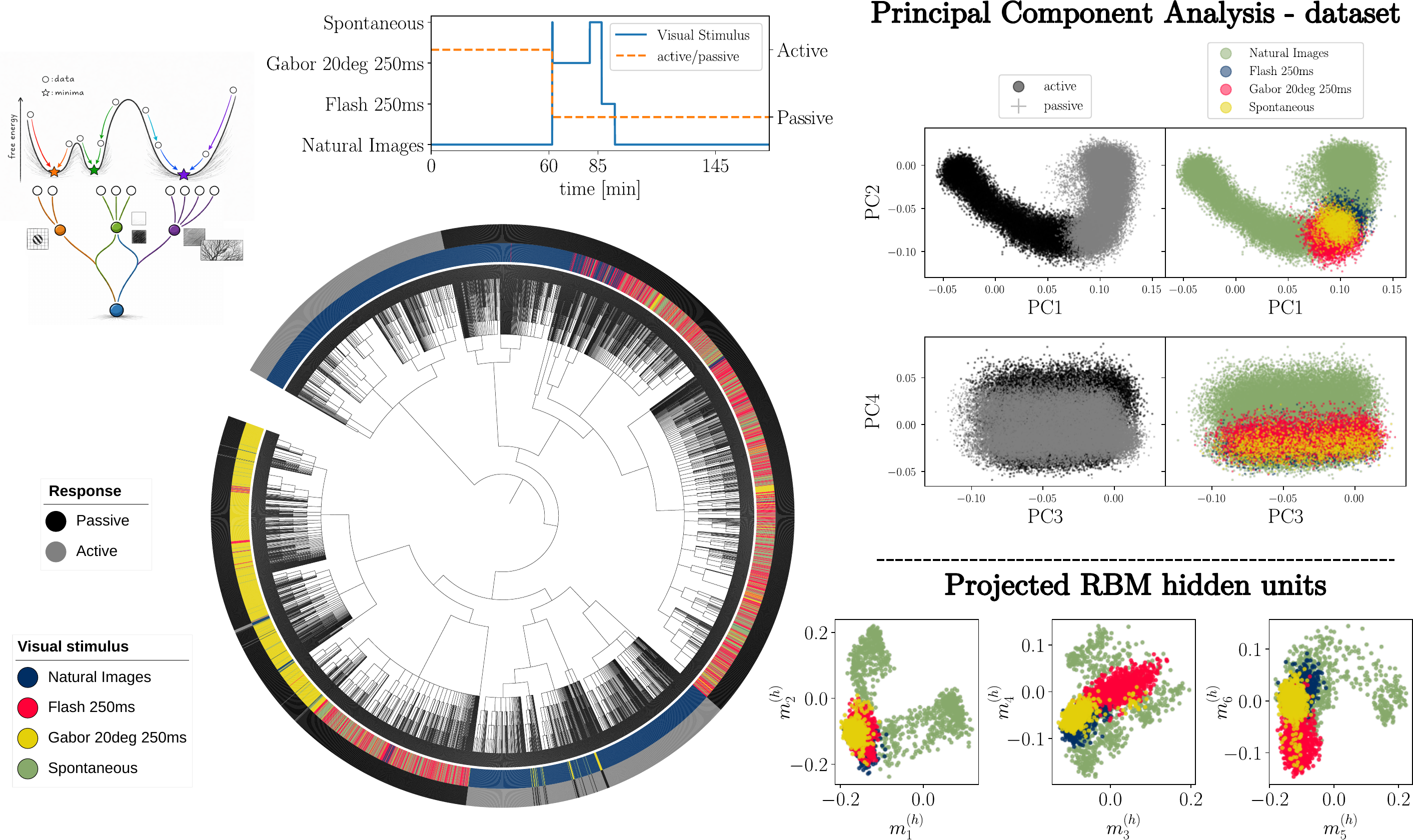}
\put(32,53){{\figpanel{a}}}
\put(63,53){{\figpanel{b}}}
\put(62,17){{\figpanel{c}}}
\put(10,35){{\figpanel{d}}}
\end{overpic}
    \caption{\textbf{Inferring activity/stimuli from the free-energy landscape.} \figpanel{a} Timeline of the experiment showing the presented visual stimuli and the mouse behavioral state during active recognition (licking task) and passive replay phases.  \figpanel{b} Neural activity projected onto the first four principal components, with data points coloured according to the label associated with each time point, illustrating that inferring the labels from this linear low-dimensional representation is a highly nontrivial task. Instead, in \figpanel{c} we project the hidden magnetizations states associated to each data-point projected on $6$ singular vector of RBM's weight matrix where projections now separate more the different colored data: for each sample $\mu$ in the dataset we compute the corresponding mean activation of the hidden units $\bm m_h^{(\mu)}=\av{\bm h}_{p(\bm h|\bm v^{(\mu)})}$ and we project the result across the SVD directions of the weight matrix. In \figpanel{d} we construct a tree from the TAP fixed points at different stages of training, using 1000 samples from each label and following the procedure of Ref.~\cite{decelle2023unsupervised}. The resulting structure shows that TAP approximations of the free-energy minima naturally cluster neural activity patterns with the same label, and are even able to correctly classify some categories despite having no access to label information during training. 
    }
    \label{fig:fixed-points}
\end{figure*}
The experimental dataset allows to access external labels for each time-dependent neural pattern recording, which are associated to either the active/passive part of the session (i.e. the mouse being performing an action in response to the change of the visual stimulus, or not) and to the type of visual stimulus itself the mouse is exposed to during the recording: Figure \ref{fig:fixed-points}--\figpanel{a} summarizes for the actual temporal evolution of each label across the experiment, with the same binning used to discretize neural activity. The question is whether the RBM identifies distinct neural activity patterns associated with different stimuli or with different behaviors performed by the mouse and, if so, whether the probability measure learned by the model can be used to cluster these patterns in an unsupervised way according to their associated stimulus or behavioral state. In Fig.~\ref{fig:fixed-points}--\figpanel{b} we show that, while the active/passive state of the mouse can already be partially inferred from the PCA representation (with neural activity patterns colored according to their labels), external stimuli are not linearly separable within this low-dimensional projection. Yet, in Fig.~\ref{fig:fixed-points}--\figpanel{c}, we show that the mean activation of the RBM hidden units, computed by fixing the visible variables to the experimental neural patterns, and projected onto the first four singular vectors of the RBM weight matrix, leads to a clearer separation of the labels, with points colored according to the label associated to each time frame.

Yet, a finer analysis can be achieved by studying how the landscape evolves during training. Once the model is trained, the RBM free-energy landscape can be analyzed using tools from spin-glass physics~\cite{gabrie2015training,maillard2019high}. The corresponding basins of attraction associated with the neural recordings provide a natural way to cluster neural activity into groups of similar patterns, while the progressive specialization of the landscape throughout training can be exploited to organize free-energy minima into family trees, as recently proposed in Ref.~\cite{decelle2023unsupervised}. This, in turn, allows neural activity recorded at different times to be clustered according to their similarity as shown in Fig.~\ref{fig:fixed-points}--(d), where active and silent periods, as well as natural and Gabor images, cluster into common branches of the tree, indicating that they are associated with the same minima of the TAP free-energy landscape.

\subsection{Simulating neural activity}
\begin{figure*}[t!]
    \centering
    \begin{overpic}[width=\textwidth]{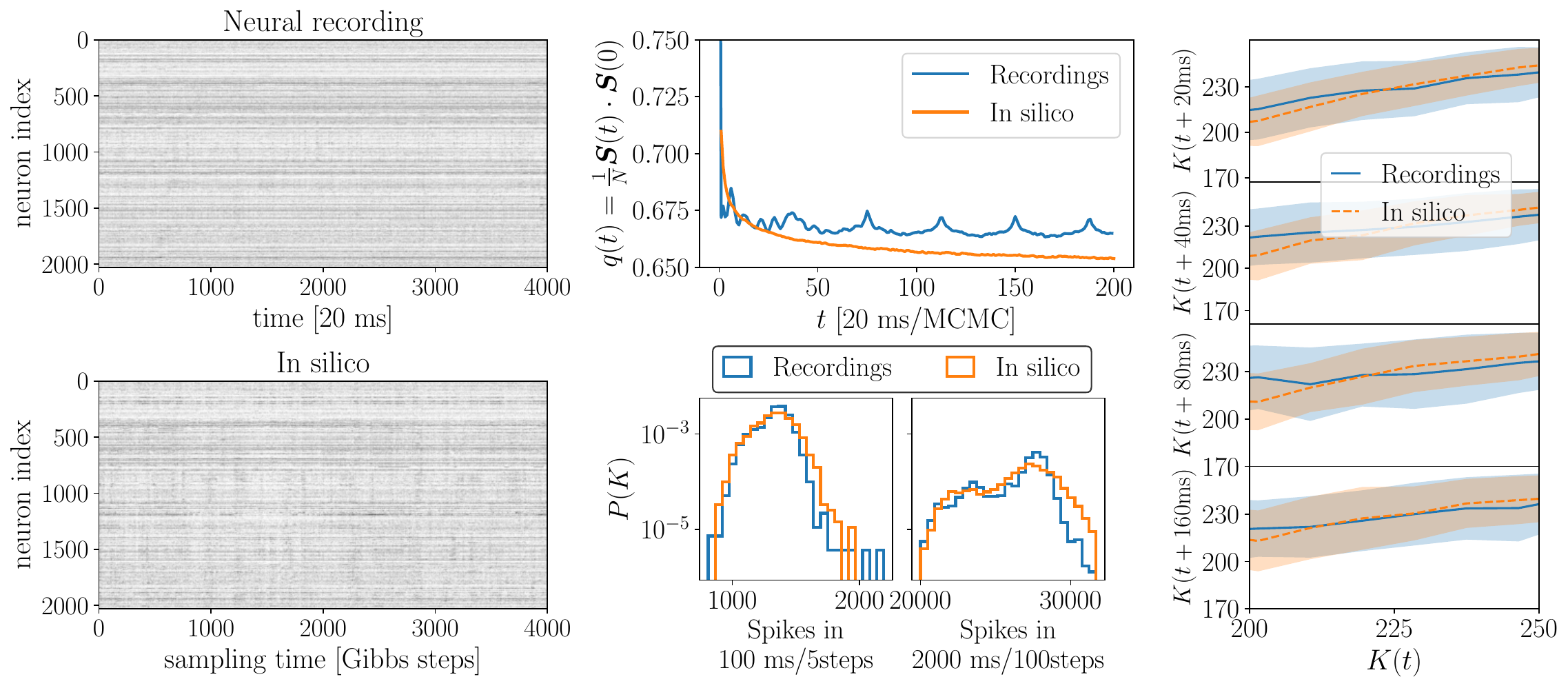}
\put(1,41){{\figpanel{a}}}
\put(1,21){{\figpanel{b}}}
\put(48,38){{\figpanel{c}}}
\put(41,18){{\figpanel{d}}}
\put(77,41){{\figpanel{e}}}
\end{overpic}
    \caption{\textbf{Generation in time.} In \figpanel{a} we show a snapshot of consecutive neuronal spikes recorded over an 80 s window. Neurons are classified as active (black) or inactive (white) using time bins of 20 ms. In \figpanel{b} we show an analogous snapshot from an in silico recording, where neural activations are generated using a MCMC process based on an RBM trained on the real neural recordings. \figpanel{c}: overlap between the neural activity $S$ at
time $T + t$ and time $T$, averaged across different starting points in the
timeseries. In \figpanel{d} we show the histogram of the number of spikes in the population measured in time windows of 100 ms (corresponding to $5$ Gibbs steps) and 2000 ms (corresponding to $100$ Gibbs steps), comparing real recordings and in silico data. In \figpanel{e} we show the mean and standard deviation of the number of active neurons $K(t+\Delta)$ conditioned on having $K(t)$ active neurons at time $t$, for $4$ values of $\Delta \in \{ 20, 40, 80, 160 \} [\mathrm{ms}]$.  The trained RBM is the same as in Fig.~\ref{fig:genstats}.
    }
    \label{fig:timedependent} 
\end{figure*}

The RBMs presented in this work are trained to reproduce the overall statistics of the data using neural responses discretized into $20$ms time bins. To construct the dataset, only one frame every $8$ is retained, because temporal correlations render intermediate frames largely redundant (see the Materials and Methods section). Formally, in our approach, we treat all samples as i.i.d. drawn from an unknown distribution that the model aims to approximate. However, the underlying recordings are time-dependent, and neural responses at different times exhibit temporal correlations. This raises the natural question of whether the learned models can provide insight into the internal dynamics of the system. In particular, one may ask how reliable the \textit{in silico} neural recordings generated through MCMC sampling are (especially when using the alternating Gibbs sampling procedure naturally associated with RBMs) and whether the resulting synthetic dynamics can capture aspects of the temporal organization of neural responses. Examples of real and synthetic recordings are shown in Figs.~\ref{fig:timedependent}-\figpanel{a} and \figpanel{b}, which display snapshots of the experimental activity and of activity generated by the RBM,  both initialized from the same experimental configuration. For the experimental recordings, each time step corresponds to $20$ ms of experimental time. For the in silico dynamics, one time step corresponds to $1$ alternating block-Gibbs step, i.e. a subsequent update of the hidden layer conditioned on the visible one (firstly) and viceversa (secondly).

Previous studies have argued that models trained on static data are unable to reproduce the temporal statistics across successive time bins~\cite{gardella2018blindfold}. Here we find that, although part of the temporal structure is indeed lost, the MCMC dynamics can nevertheless capture the overall relaxation of the system. This is clear from Fig.~\ref{fig:timedependent}-\figpanel{c} where we show the overlap between the neural activity at time $T+t$ and the same quantity computed at time $T$
(and the result is averaged across different points in the timeseries): the models fail to reproduce the oscillatory behavior in the recordings, but still qualitative captures the overall relaxation.
Notably, the synthetic dynamics captures several coarse-grained temporal statistics of the neural activity: in particular, it reproduces the distribution of the total number of spike counts computed over larger temporal windows, such as $100$ ms or $2000$ ms bins (Fig.~\ref{fig:timedependent}-\figpanel{d}), as well as the conditional statistics of the population activity, measured through the average number of active neurons $K(t+\Delta)$ given the number of active neurons $K(t)$ at time $t$. This quantity is shown in Fig.~\ref{fig:timedependent}-\figpanel{e} for several values of $\Delta$, where the line represents the mean and the shaded region the standard deviation, again computed across different starting points $T$. \\ Finally, we emphasize that this dataset is not ideally suited for this comparison because the mice are continuously driven by external sensory stimuli. Since these inputs are not provided to the model, it cannot reproduce the exact temporal evolution of the recordings but only their underlying statistical structure. Consequently, the comparison at long times is intrinsically unfair.

\section{Conclusions}
In this work we demonstrate that Restricted Boltzmann Machines provide a powerful and scalable framework for modeling large-scale neural population activity. Using Neuropixels recordings as datasets, we show that RBMs accurately reproduce the empirical statistics of the data while maintaining a compact and interpretable parametrization. In particular, the trained models match first-, second-, and higher-order statistical moments of the neural activity, as well as global observables such as the distribution of population activity and the energy landscape of the inferred model. Thanks to a controlled training procedure and on-line monitoring of the log-likelihood, it is possible to select models that best describe a given dataset while minimizing overfitting.

Beyond their generative capabilities, RBMs provide a principled route to extracting interpretable information about neural interactions. By mapping the trained model onto an effective multibody interaction representation, we infer fields and interaction terms that reveal statistical dependencies among neurons. The resulting interaction patterns are heterogeneous and sparse, with most couplings being weak and only a small subset contributing to substantial interactions. Importantly, the inferred couplings reflect known aspects of brain organization: neurons in visual cortical areas exhibit stronger interactions among themselves, consistent with their coordinated activity during visually driven tasks. We can also quantify the importance of higher-than-two-body interactions, which shows large variations across brain areas.
These results illustrate how RBMs bridge expressive generative modeling with the statistical–mechanical interpretation traditionally pursued in maximum-entropy approaches.

Although the models were trained on temporally shuffled snapshots of neural activity, we find that the dynamics induced by the RBM sampling procedure reproduces several coarse-grained temporal statistics of the recordings, including the relaxation of population activity and the distribution of spike counts over extended time windows. While this synthetic dynamics does not capture all temporal features—such as oscillatory structures in autocorrelation functions—it nevertheless suggests that generative models trained on static statistics can still encode aspects of the system’s dynamical responses.

Several directions for future work naturally emerge from this study. First, extending the present framework to explicitly incorporate temporal dependencies—for instance through temporal or recurrent variants of RBMs \cite{temporal_RBM_hinton,gardella2018blindfold,monnens24the}—could enable a more faithful representation of neural dynamics and stimulus-driven responses. Second, the latent variables learned by the model provide a low-dimensional representation of neural population activity whose structure deserves further investigation, particularly in relation to behavioral variables, stimulus features, or internal brain states. Third, comparing models trained under different behavioral conditions (such as familiar versus novel stimuli) may reveal how effective interaction networks reorganize during learning or plasticity. Finally, the generative nature of the RBM framework enables controlled in silico perturbations—such as clamping or modifying subsets of neurons—to probe causal relationships within neural populations.

Taken together, these results establish Restricted Boltzmann Machines as grounded statistical-physics–inspired approach for analyzing large-scale neural recordings, and for systematic testing of predictive hypotheses linking the statistical structure of the neural activity to behavioral and experimental variables. By combining scalability, generative power, and interpretability, RBMs provide a useful tool to uncover the collective structure of neural activity and to connect high-dimensional experimental data with effective models of interacting neuronal populations.

\section*{Materials and Methods}

\subsection{Training of Restricted Boltzmann Machines}
Training an RBM amounts to maximizing the log-likelihood $\mathcal{L}(\bm{\theta}|\mathcal{D})$ of the observed data $\mathcal{D}=\{\bm v^{(m)}\}_{m=1}^M$ w.r.t the parameters $\bm{\theta}$, which is equivalent to minimizing the Kullback--Leibler divergence between the empirical data distribution and the model distribution:
\begin{equation}
\mathcal{L}(\bm{\theta}\mid \mathcal{D})
=
\left\langle
\log p_{\bm{\theta}}(\bm{v})
\right\rangle_{\mathrm{data}} . \label{eq:lldef}
\end{equation}
Here $p_{\bm{\theta}}(\bm{v})$ denotes the marginal distribution over visible configurations, given by \eqref{eq:pmarginal}.
The gradient of the log-likelihood with respect to a parameter $\theta_\alpha \in \{W_{ia}, b_i, c_a\}_{i=1,\ldots,\Nv}^{a=1,\ldots,\Nh}$ can be written in the generic form
\begin{equation}
\frac{\partial \log \mathcal{L}}{\partial \theta_\alpha}
=
\left\langle
\partial_{\theta_\alpha} \mathcal{H}
\right\rangle_{\mathrm{data}}
-
\left\langle
\partial_{\theta_\alpha} \mathcal{H}
\right\rangle_{p_{\bm{\theta}}},
\label{eq:ll_grad}
\end{equation}
where $\left\langle \cdot \right\rangle_{\mathrm{data}}$ denotes the average over the empirical distribution
$
p_{\mathrm{data}}(\bm v)
=
\frac{1}{M}
\sum_{m=1}^{M}
\delta\!\left(\bm v-\bm v^{(m)}\right),
$
and $\left\langle \cdot \right\rangle_{p_{\bm{\theta}}}$ denotes the expectation with respect to the model 's marginal distribution $p_{\bm{\theta}}(\bm v)$ defined in \eqref{eq:pmarginal}. Each contribution to the gradient therefore represents the difference between equilibrium expectation values under the model distribution and empirical averages measured on the data, evaluated for the observable conjugate to $\theta_\alpha$.

From a statistical physics perspective, training an RBM means tuning local fields and interaction strengths so that its equilibrium distribution~\eqref{eq:pmarginal} reproduces, as accurately as possible, the empirical statistics of the data: similarly to maximum entropy formulations, log-likelihood maximization imposes a set of moment-matching conditions—equal in number to the model parameters—requiring that the expectation values of the observables $\bm{O} = \nabla_{\bm{\theta}} \mathcal{H}$ coincide under the measures $p_{\mathrm{data}}$ and $p_{\bm{\theta}}$\cite{agoritsas2023explaining}. There is however an important difference to highlight: as opposite to maximum entropy approaches, where a specific set of observables is externally selected to constrain the model -- e.g. single-site and pairwise correlations -- here the observables $\bm{O}$ that are constrained come naturally from the marginalization over visible nodes of the RBM: a simple calculations shows indeed that these observables can be expressed as combination of all moments in the data, as a result of the non-linear effect of hidden variables into the marginalization.

The main caveat of this approach lies in the intractability of the partition function $Z_{\bm \theta}$, which prevents an exact evaluation of the gradient. The data-dependent contribution (i.e., the first term in \eqref{eq:ll_grad}) is straightforward to compute, as it reduces to empirical averages over the training dataset. In contrast, the model-dependent expectation $\left\langle \cdot \right\rangle_{p_{\bm{\theta}}}$ involves an exhaustive sum over all configurations of visible and hidden variables weighted by the Boltzmann distribution, whose computational cost grows exponentially with system size. 
In practice, this contribution is estimated using MCMC methods. Recent works have emphasized the importance of controlling the convergence of the sampling process during training to avoid non-equilibrium effects and obtain reliable models~\cite{decelle2021equilibrium,agoritsas2023explaining}. In this work, we employ recent advances in RBM sampling, training, and evaluation based on Trajectory Parallel Tempering (PTT)~\cite{bereux2024fasttrainingsamplingrestricted}. Specifically, we use the algorithm and implementation recently introduced in~\cite{bereux2026ptt} to estimate the negative phase of the likelihood gradient. This approach guarantees MCMC convergence throughout training, dynamically adjust the learning rate, and significantly improves sampling quality and training stability over previous methods~\cite{bereux2026ptt}, while maintaining a computational cost comparable to that of standard training algorithms. The same PTT framework is also used to generate samples from the trained models and to compute the log-likelihood along the training trajectory. All sampling procedures used for sample generation and log-likelihood estimation are systematically verified for thermalization using the protocol described in~\cite{bereux2026ptt}.

\subsection{Extraction of effective model}
Marginalization over the hidden units yields an explicitly energy-based model for the data distribution, allowing RBMs to learn complex probability landscapes. Starting from the marginal distribution over visible nodes Eq.~\ref{eq:energy_visible}, and introducing spin variables $s_i \!\in\! \{ -1, +1 \}$ defined by $s_i \equiv 2 v_i - 1$, it has been recently shown that the marginalized energy of the RBM can be exactly expanded as an Ising-like model with higher-order interactions~\cite{decelle2024inferring},
\begin{equation}
   \!\!\! \mathcal{H}(\boldsymbol{s}) = - \sum_i H_i s_i - \sum_{i < j} J_{ij}^{(2)} s_i s_j - \sum_{i < j<k} J_{ijk}^{(3)} s_i s_j s_k  + \dots,\!\!\!
   \label{eq:Hamiltonian_effectivemodel}
\end{equation}
with effective fields $H_i$ and couplings $J_{i_1 \dots i_n}^{(n)}$ given by
\begin{align}
    H_i 
    &= \frac{1}{2^{\Nv}} \sum_{\boldsymbol{s}'} \sum_a {s}'_i \ln \cosh \left( \sum_j w_{ja} s'_j + \zeta_a \right) + \eta_i, \label{eq:fields_formula}\\
    J_{i_1 \dots i_n}^{(n)} 
    &= \frac{1}{2^{\Nv}} \sum_{\boldsymbol{s}'} \sum_a \prod_{k=1}^n s'_{i_k} \ln \cosh \left( \sum_j w_{ja} s'_j + \zeta_a \right). \label{eq:couplings_formula}
\end{align}
Here, we have introduced a reparameterization of the model defined by $ \eta_i \equiv \frac{1}{2} \left( b_i + \frac{1}{2} \sum_a w_{ia} \right), \ \theta_i \equiv \frac{1}{2} \left( c_i + \frac{1}{2}  \sum_i w_{ia} \right) , \ \mathrm{and} \ w_{ia} \equiv \frac{1}{4} W_{ia}$. 
In \eqref{eq:couplings_formula}, the evaluation of each coupling requires a sum over $2^{N_v}$ configurations, which is computationally prohibitive. To circumvent this limitation, we employ a Gaussian approximation introduced in Ref.~\cite{decelle2024inferring}, which makes the computation tractable and naturally parallelizable. Since most of the coupling weight is typically concentrated in low--order terms, computing all higher--order couplings is not necessary to obtain a reliable representation of the interactions between variables. Additionally, to identify which pairs of variables participate in non-negligible interactions beyond second order, we introduce the following log--likelihood ratio 
\begin{equation}
r_{ij}^{(2)}(\boldsymbol{s'}) = \frac{1}{4} \sum_a \!\sum_{s_i, s_j } \! s_i s_j  \ln \cosh \Big( w_{ia} s_i + w_{ja} s_j + \!\!\!\sum_{k\notin \{ i, j \} } \!\!\! w_{ka} s'_k + \zeta_a  \Big).
\label{eq:log-likelihoo_ratio}
\end{equation}
Note that when sites $i$ and $j$ interact purely through pairwise couplings, \eqref{eq:log-likelihoo_ratio} becomes independent of $\boldsymbol{s'}$ and reduces to $r_{ij}^{(2)}(\boldsymbol{s'}) = J_{ij}$. Therefore, the presence of higher--order interactions can be quantified through the fluctuation
\begin{equation}
    R_{ij}^{(2)} = \sqrt{ \frac{1}{2^{\Nv}} \sum_{\boldsymbol{s'}} \left( J_{ij}^{(2)} - r_{ij}^{(2)}(\boldsymbol{s'})\right)^2 } \label{eq:R2_def}
\end{equation}
which we estimate through Monte Carlo integration.
The observable $R^{\paren{2}}_{ij}$ allows to compute an estimate of whether high-order interaction terms are to be expected among any $n-$tuple of variables involving both $i$ and $j$.

As a final note, we stress again that in the RBM the interactions appearing in its effective Ising model formulation are not independent parameters to be fit like in the maximum entropy construction, but rather they all depend on the architectural parameters of the RBM $\bm \theta $. 

\subsection{Information gain\label{sec:MMinformationgain}} Following~\cite{carcamo2025statistical}, we quantify the information gain as $I_{\rm G} = S_{\mathrm{ind}} - S_{\mathrm{RBM}}$, where $S_{\mathrm{ind}}$ is the entropy of an independent model fitting only the local neural activities (i.e., the first moments), and $S_{\mathrm{RBM}}$ is the entropy of the trained RBM. The latter can be computed from the training log-likelihood as
\begin{equation}
S_{\text{RBM}}= \left\langle \mathcal{H}\left(\boldsymbol{v}\right)\right\rangle_{p_{\boldsymbol{\theta}}} -\left\langle \mathcal{H}\left(\boldsymbol{v}\right)\right\rangle _{\text{data}}-\mathcal{L}
\end{equation}
where $\mathcal{L}$ is the log-likelihood of the training set (from ~\eqref{eq:lldef}) computed at the early-stopping point, $\mathcal{H}\left(\boldsymbol{v}\right)$ is the energy of visible configurations defined in ~\eqref{eq:energy_visible} and the first average is computed using a set of thermalized chains. The result in Fig.~\ref{fig:comparison_rats}-\figpanel{c} is further normalized by the number of neurons $N$.

\subsection{Data pre-processing and details of training procedure}
All the data used in this work are publicly available in the Allen Institute of Brain Science \cite{AllenData}. Each recording session corresponds to a unique experiment where a mouse performs a change-detection task driven by visual stimuli, followed by a passive interval where different visual stimuli are presented; during all the experiment, neural activity is recorded through NeuroPixels probes. Each experimental session has a duration of about $3$ hours and comprises several stimulus blocks, including natural images and Gabor stimuli (grating patches), with stimulus presentations in windows of $\sim 250$ms. 
Each mouse is recorded on two consecutive days, one session with a familiar image set (used during training) and one with a novel image set, enabling analyses of experience-related differences in neural coding. 
 
The raw data contain absolute (real) firing times of each neuron (obtained after spike-sorting): therefore, for our purposes a discretization is needed to extract single-time frames where each of them will account for the collective activity of a population of neurons within that window. A time-binning $\Delta= 20$ms is used. Within each time slice, all neurons which are recorded having at least one firing signal in the interval $\caja{t, t+ \Delta}$ are assigned a $+1$ value, and $0$ otherwise. This procedure generates a dataset of binary entries $\lazo{0,1}$ composed by $T \, / \Delta$ time frames, where $T$ is the total experimental window chosen. All the experimental session except the last half-hour is chosen. The initial set of time frames is further reduced by a factor $8$ to reduce the dataset's size. Data are then shuffled and split in a train and test set, with $0.7/0.3$ proportion.
Additional details about the experimental setup, data acquisition, post-processing is available at~\cite{VisualBehavior_whitepaper}. The results shown in the main text refer to  experimental sessions where the chosen mouse is ``familiar'' with the visual stimulus. We randomly select 8 experimental sessions /mice among the  $54$ available, with a number of stable recorded neurons in the range $\sim 1500-2200$ (a full table of values is given in the SI).

\subsection{Model selection} The main architectural parameter in a RBM is the number of hidden nodes $\Nh$, which tunes the expressivity of the model in taking into account the dataset's correlations.
The intuitive picture we expect for a trained RBM with a given number of hidden nodes is that, when $\Nh$ is too low, the model will not be able to fit all the non-trivial correlations in the observed data, making it poorly expressive. On the other hand, if $\Nh$ grows too much, the model has too many parameters for the dataset that needs to be learned, with consequent problems of overfitting.
In order to choose the minimal model that captures all the complexity in the data but without overfitting, we perform model selection by looking at the log-likelihood computed on a test set along the training trajectory. The log-likelihood on a test set indicates how well the model is able to generate data that has never seen, making it a reliable criterion to measure the model’s ability to generalize beyond the training set.
However, computing the log-likelihood for an RBM is non-trivial, since it requires estimating the partition function $Z_{\bm \theta}$ in \eqref{eq:BoltzmannDistributionRBM}, which is generally intractable and must be approximated using specialized techniques. 
Standard procedures involve e.g. annealed importance sampling (AIS) techniques, which however do not guarantee a reliable evaluation if not in the limit of an infinitesimal temperature annealing protocol. In this work, we exploit the parallel trajectory tempering (PTT) technique introduced in~\cite{bereux2024fasttrainingsamplingrestricted}.  In practice, as model capacity or training time increases, the training log-likelihood may continue to improve while the test log-likelihood eventually reaches a maximum and then begins to deteriorate, signaling the onset of overfitting. Therefore, the optimal model is selected at the point where the test log-likelihood is maximized, prior to the regime in which overfitting dominates.

\section*{Acknowledgments}
The authors acknowledge financial support from grant PID2024-158623NB-C21, funded by MICIU/AEI/10.13039/501100011033 and by ERDF/EU.  This work is supported by the European Research Council (ERC) under the European Union's Horizon Europe research and innovation programme (Grant Agreement No. 101231393 — BeME).
FM was supported by fellowships from the Simons Foundation and the Human Frontier Science Program. G.C. acknowledges funding from the Spanish Ministerio de Ciencia, Innovación y Universidades through project María de Maeztu CEX2021-001164-M funded by the MICIU/AEI/10.13039/501100011033.

\bibliographystyle{apsrev4-2}
\bibliography{sample}

\appendix
\onecolumngrid
\newpage

\section{Model selection}
\begin{figure}[h!]
    \centering
\begin{overpic}[width=0.8\textwidth]{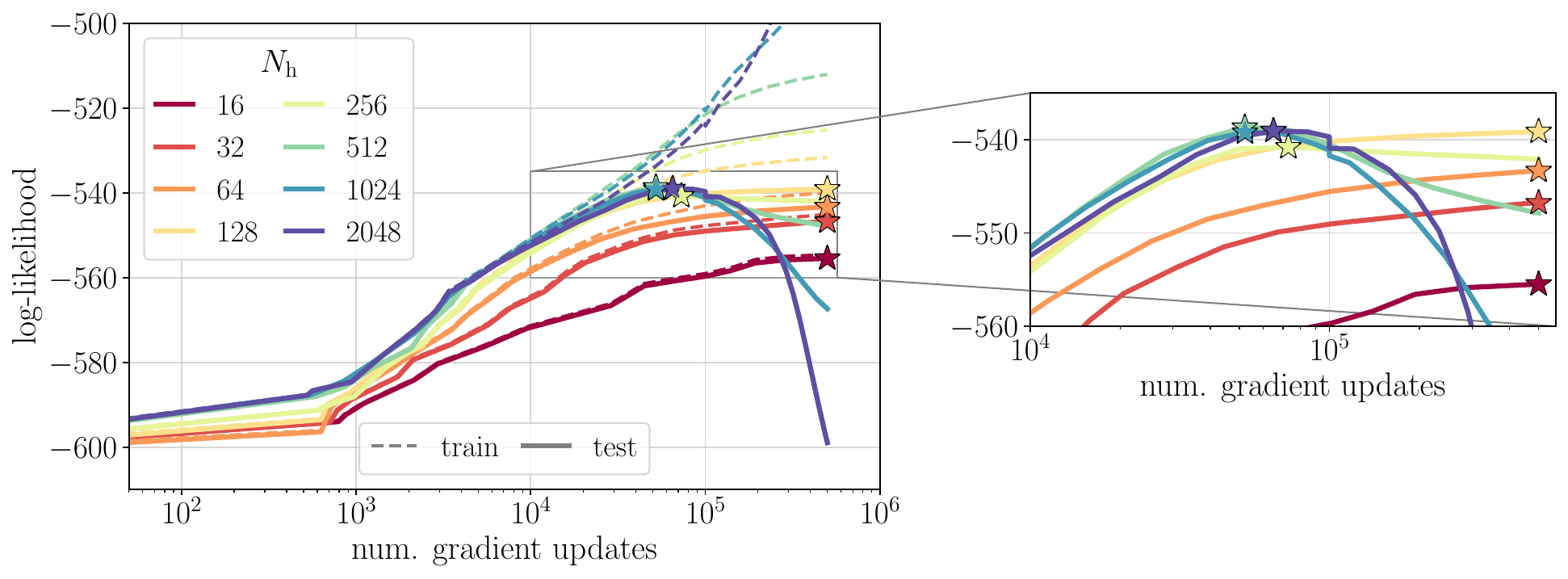}
\end{overpic}
    \caption{\textbf{Model selection}. We plot the evolution of the train and test log-likelihood (resp.\ in dashed and solid lines) along the training trajectory, for different trainings performed by varying the number of hidden nodes $\Nh$ on a single dataset (corresponding to mouse ID 560771). For each curve the star symbol indicates the update at which the maximum test log-likelihood is reached. }
    \label{fig:ll_varying_Nh}
\end{figure}

We report here a comparison of different trainings with different amounts of hidden nodes $\Nh$, to select the optimal model according to what discussed in Section 3.4 (see the Materials and Methods section) in the main text. The comparison is shown in Fig~\ref{fig:ll_varying_Nh}, where we plot  the evolution of the train (resp.\ test) log-likelihood in dashed (resp.\ solid) lines vs.\ the training time (i.e.\ the number of gradient updates). All machines are trained on the same dataset with the same protocol discussed in the Materials and Methods section of the main text.  We can observe how at small $\Nh$ in the range $\Nh \in \paren{2,128}$ the test log-likelihood saturates at a plateau (with increasing height as $\Nh$ increases): therefore, increasing $\Nh$ and the number of RBM's weights increases the model's expressivity. On the contrary, for $\Nh=256,512, 1024,2048$ the test log-likelihood first reaches a maximum and then starts to decrease, with a steeper behavior for increasing $\Nh$ (as clear also from the zoomed inset at the right). This is a clear signal that the RBM starts to overfit the training data from this point onward (the maximum test log-likelihood point is highlighted with a star symbol in each curve). 

\begin{figure}[h!]
    \centering
\begin{overpic}[width=\textwidth]{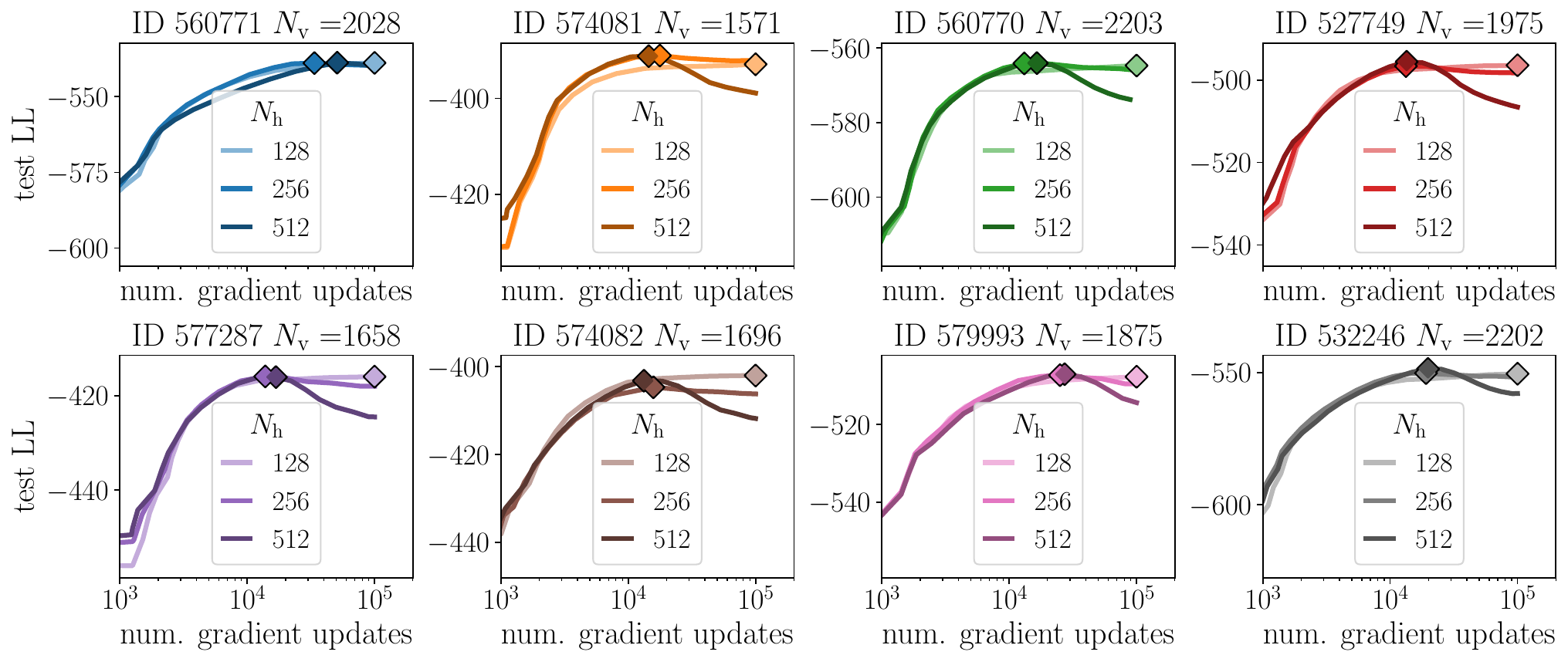}
\end{overpic}
    \caption{ \label{fig:ll_varying_mouse} \textbf{Model selection}. We plot the evolution of the train and test log-likelihood (resp.\ in dashed and solid lines) along the training trajectory, for different trainings performed by varying the number of hidden nodes $\Nh$ in a smaller range than in Fig.~\ref{fig:ll_varying_Nh}, for $8$ datasets associated to $8$ different mouses (one for each panel). For each curve the diamond symbols indicate the update at which the maximum test log-likelihood is reached.}
\end{figure}

We further report a model selection comparison (again, based on the test-log likelihood) for the full set of $8$ experimental sessions / mice considered in this work. The results are shown in Fig.~\ref{fig:ll_varying_mouse}, where for each panel we performed $3$ different trainings with $128,256,512$ hidden units respectively. We can observe how the log-likelihood behaves qualitatively in the same way across experiments, with, and that the number of hidden nodes needed to best describe these data sits in the range $\Nh \in [128,256]$ with a significant robustness against the specific animal chosen. For this reason, in the  main text we adopt the overall optimal value of $\Nh=256$ hidden nodes for all the experimental sessions.

With this value, the total number of parameters to be learnt by the RBM for e.g. the mouse ID 560771 (the one considered in Figures 1, 5 , 6  in the main text) is approximately $\Nv \times \Nh \sim 5 \times 10^5$. By comparison, a fully connected Boltzmann machine trained on the same data to match only first- and second-order statistics would require fitting $\binom{\Nv}{2} \sim 2 \times 10^6$ independent pairwise interactions. Thus, with this choice of $\Nh$, the RBM architecture uses roughly one fourth as many parameters as the BM while offering greater expressivity for modeling higher-order correlations. Moreover, RBMs are much faster to train, since block-Gibbs sampling allows simultaneous updates of all visible variables, whereas BMs require sequential updates.

\section{Additional computational details}

This section contains additional computational details and numerical procedures that are not
fully specified in the main text to reproduce results.
 
\subsubsection{RBM training hyperparameters}

\begin{itemize}
\item[---] \textbf{Training protocol \& Model selection}: We train all RBMs using the PTT training protocol and implementation of Ref.~\cite{bereux2026ptt}. The method relies on PTT MCMC sampling algorithm~\cite{bereux2024fasttrainingsamplingrestricted} to estimate the negative term of the gradient and adaptively adjusts the learning rates from the cosine similarity measure between successive gradient updates, and using separate learning rates for the weights and the visible and hidden biases. The weights are initialized as i.i.d. Gaussian variables with variance $10^{-8}$, the hidden biases are set to zero, and the visible biases are initialized as
\begin{equation}
b_i=\log \av{\sigma_i}_{\mathrm{data}}
-\log\!\left(1-\av{\sigma_i}_{\mathrm{data}}\right),
\end{equation}
so that the zero-weight RBM reproduces the empirical first moments. 

We use $1000$ parallel chains and a minibatch size of $1000$, with maximum learning rate fixed to $0.1$. Trainings are run for $10^5$ gradient updates for all mice, except mouse ID 56077, for which $10^6$ updates are used to make the comparison in Fig.~\ref{fig:ll_varying_Nh} because models with fewer hidden units tend to converge slower. In all the trainings, no regularization is applied: overfitting is controlled by a posteriori model selection and early stopping, choosing the model with the highest test log-likelihood and, among models with comparable performance, the smallest number of parameters, as illustrated in Figs.~\ref{fig:ll_varying_Nh} and~\ref{fig:ll_varying_mouse}. Further details are given in Ref.~\cite{bereux2026ptt}.
\item[---] \textbf{Sample generation protocol.} We generate samples using the PTT sampling algorithm~\cite{bereux2024fasttrainingsamplingrestricted}, and systematically check for thermalization of the chains before computing any analysis. We follow the thermalization protocol used in~\cite{bereux2026ptt}, monitoring the diffusion of replicas along the PTT ladder through the autocorrelation of their checkpoint indices. Its long-time exponential decay defines the relaxation time $\tau_{\mathrm{exp}}$, and each PTT ladder is evolved for at least $20\tau_{\mathrm{exp}}$ before sampling. After thermalization, configurations are recorded every $2\tau_{\mathrm{int}}$, where $\tau_{\mathrm{int}}$ is the corresponding integrated autocorrelation time.
\item[---] {\bf Log-likelihood estimation.} We estimate the log-likelihood using the equilibrium PTT-based method of Ref.~\cite{bereux2026ptt}, after verifying thermalization of the chains. We further validate these estimates against Annealed Importance Sampling (AIS) using sufficiently long chains and dense temperature ladders, finding consistent results.
\end{itemize}

\subsubsection{Training times}
Table~\ref{tab:tabletimes} reports the wall-clock time required to train an RBM on an NVIDIA RTX 3090 GPU for each mouse dataset, using $N_\mathrm{h}=256$. We further report the dataset's size (i.e. the number of visible nodes $\Nv$) for completeness. In all cases, we limited the maximum learning time to $0.01$ for the adjustable learning-rate scheduler.
\begin{table}[]
    \centering
    \setlength{\tabcolsep}{3pt} 
    \begin{tabular}{ccc}
         \textbf{Mouse ID} & $\boldsymbol{N_\mathrm{v}}$  & \textbf{time PTT} \\\hline
         560771 & 2028 & 1778.33 s \\
         574081 & 1571 & 1916.31 s\\
         527749 & 1975 & 2157.11 s\\
         532246 & 2202 & 2445.87 s\\
         560770 & 2203 & 2176.15 s\\
         574082 & 1696 & 1651.76 s\\
         577287 & 1658 & 1632.54 s\\
         579993 & 1875 & 1919.51 s\\
    \end{tabular}
    \caption{Number of neurons and wall-clock time required to train the RBM for $10^5$ parameter updates with $\Nh=256$ hidden nodes on an NVIDIA RTX 3090 GPU with the PTT algorithm described in Ref.~\cite{bereux2026ptt}.}
    \label{tab:tabletimes}
\end{table}

\subsubsection{On the statistics computation}

\begin{itemize}
          \item[---] \textbf{Three-body connected correlations.} With $N\approx
        2000$ there are ${\sim}10^{9}$ triplets to compute 3-body correlations on. After computing them, Figure~2-- \figpanel{e} in the main text only shows the union of the $M=10^5$ strongest correlations among the three sets (train, test and generated), resulting in $\approx 1.3 \cdot 10^5$ triplets. This procedure ensures the eventual inclusion of strong false positive/ false negatives (e.g. between generated samples and training data) in the triplet set.

  \item[---] \textbf{Correlation between pairs of brain areas.} To generate the plot in Fig.~2-- \figpanel{f} of the main text, we compute the correlation matrix $C_{AB} = \langle Y_A Y_B \rangle -  \langle Y_A \rangle \langle Y_B \rangle$ where 
  \begin{equation}
      Y_A = \frac{1}{n_A}\sum_{i \in A} x_i
  \end{equation} where $x_i$ is the (binarized) activity of neuron $i$ and $n_A$ is the number of recorded neurons in region $A$. The average $\langle \cdot \rangle$ is then performed either on the joint set of train and test data or on the set of generated samples. 
\end{itemize}
 \subsubsection{Effective-coupling extraction numerics (Sec.~3.2)}
 
\begin{itemize}
  \item[---] The Ising mapping to reconstruct microscopic interactions requires a Gaussian integral that approximates the sum over $\boldsymbol{s'}$ in Eqs. 12-13 of the main text. In this case, we used $100$ integration points with $\sigma=5$ standard deviations of integration bound. 
  \item[---] The computation of $R_{ij}^{(2)}$ is done using $1000$ samples for each experiment. 
  \item[---] The normalized Frobenius norm shown in Fig.~3-\figpanel{d} in the main text are computed as
  \begin{equation}
    \mathcal{F}^{\paren{n}} = \sqrt{\frac{1}{\binom{\Nv}{n}} \sum_{i_1 < \ldots < i_n} \paren{J^{\paren{n}}_{i_1, \ldots,i_n}}^2} \qquad \forall n \in \{1,2,3\}
    \label{eq:frobenius}
\end{equation}
where $n$ is the order of the inferred couplings.
\end{itemize}

\section{Supplementary figures}

In the following, we report another set of result analogous to the ones presented in the main text for another dataset, corresponding to another experimental session on a different mouse in the same conditions as before. This time, the available recorded neurons are $\Nv=1571$ (corresponding to the second column in Fig.~3 of the main text). 
The results are all qualitatively similar to the ones shown in the main text, confirming robustness of the RBM-based inference and training. 

\begin{figure*}[h!]
    \centering
    \begin{overpic}[width=\textwidth]{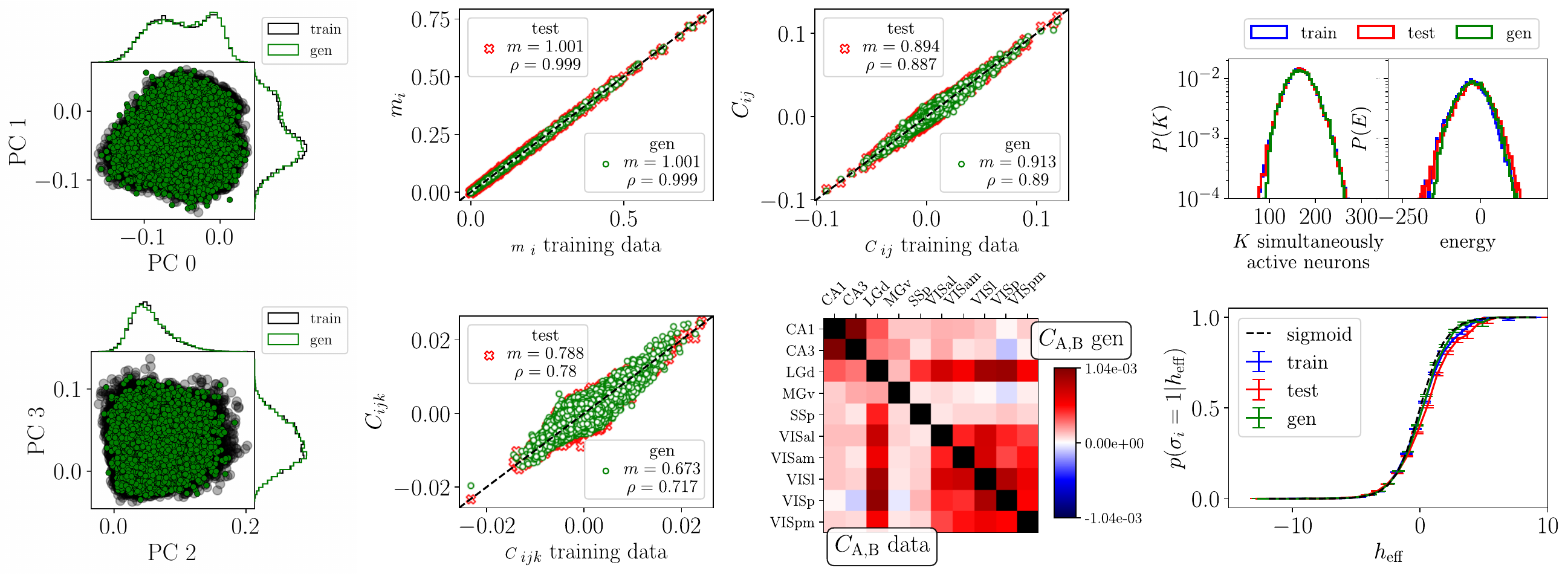}
    \put(1,33){{\figpanel{a}}}
    \put(1,15){{\figpanel{b}}}
    \put(24,35){{\figpanel{c}}}
    \put(47,35){{\figpanel{d}}}
    \put(24,15){{\figpanel{e}}}
    \put(47.5,15){{\figpanel{f}}}
    \put(75,34){{\figpanel{g}}}
    \put(95,30){{\figpanel{h}}}
    \put(75,18){{\figpanel{i}}}
\end{overpic}
    \caption{
    \textbf{Generated versus real data statistics.} Quality of generated samples assessed by comparing empirical statistics of $N=1571$ neurons with those of the training and test sets, using $10{,}000$ randomly selected sequences from each of the training, test, and generated datasets. Panel \figpanel{a}: projection of training data (in black) and generated samples (in red) along the first two principal components (PCA directions) of the training set of data;  histograms at the top and at the right highlight the one-dimensional marginal distributions over the first and second direction, respectively. \figpanel{b}: same projection but on the 2-dimensional space spanned by the third and fourth PCA directions of the data. Panel \figpanel{c}: single-neuron firing rate; \figpanel{d}: neuron-neuron covariance matrix; \figpanel{e}: three-body correlations. Red points denote training–test comparisons; green points denote training–generated comparisons. The slope of a linear fit and the Pearson correlation coefficient are reported in the legend. Panel \figpanel{f}: coarse grained connectivity between brain areas, comparison between training data (left-lower triangular part) and generated data (right-upper triangular part).
Panel \figpanel{g}: histogram of the number of simultaneously active neurons; \figpanel{h}: histogram of sample energies assigned by the RBM; \figpanel{i}: probability that a neuron is active conditioned on the network state: points are obtained by averaging over bins with similar values of $h_\mathrm{eff}$ (defined in Eq.~6 of the main text); error bars correspond to the standard error of the mean.
    }
    \label{fig:genstats_SI}
\end{figure*}
\begin{figure*}[h!]
    \centering
    \begin{overpic}[width=.9\textwidth]{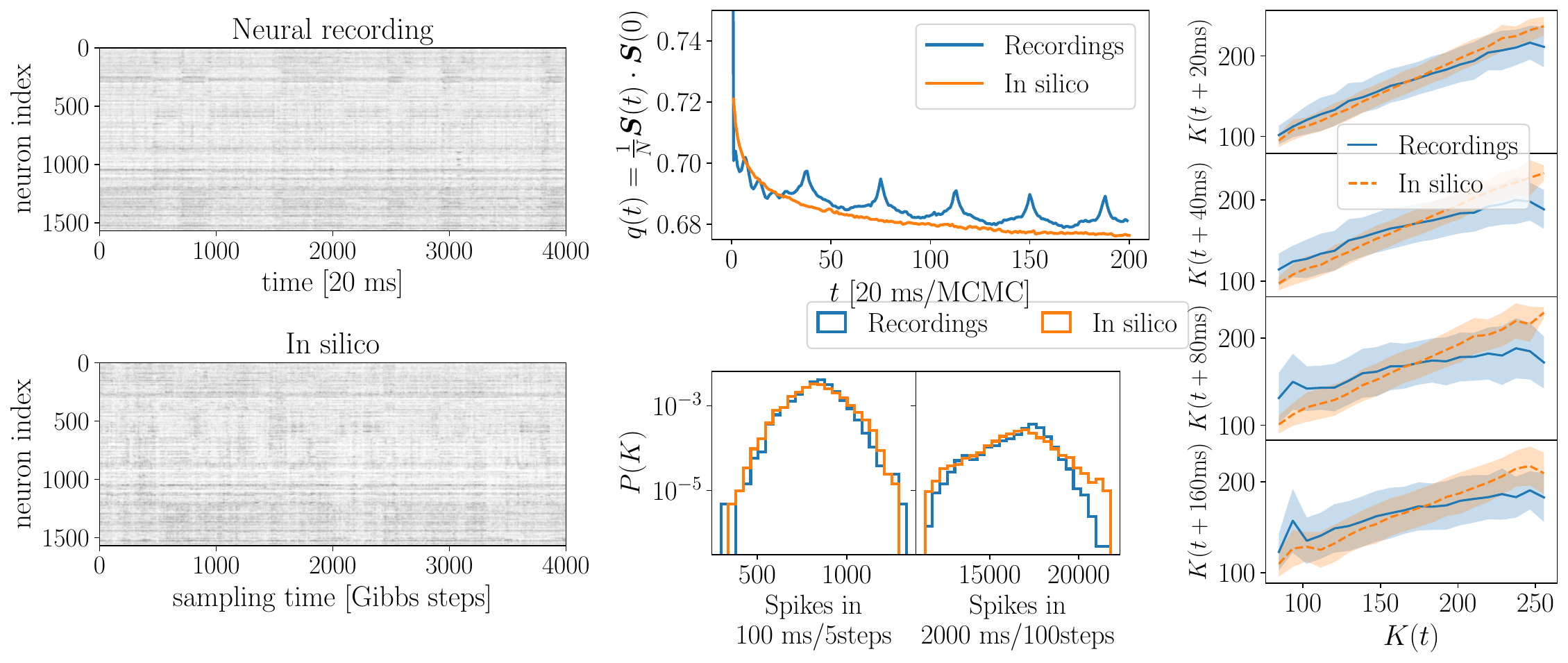}
\put(1,41){{\figpanel{a}}}
\put(1,21){{\figpanel{b}}}
\put(48,38){{\figpanel{c}}}
\put(41,18){{\figpanel{d}}}
\put(77,42){{\figpanel{e}}}
\end{overpic}
    \caption{\textbf{Generation in time.} In \figpanel{a} we show a snapshot of consecutive neuronal spikes recorded over an 80 s window. Neurons are classified as active (black) or inactive (white) using time bins of 20 ms. In \figpanel{b} we show an analogous snapshot from an in silico recording, where neural activations are generated using a MCMC process based on an RBM trained on the real neural recordings. In \figpanel{d} we show the histogram of the number of spikes in the population measured in time windows of 100 ms (corresponding to $5$ Gibbs steps) and and 2000 ms (corresponding to $100$ Gibbs steps), comparing real recordings and in silico data. In \figpanel{e} we show the mean and standard deviation of the number of active neurons $K(t+\Delta)$ conditioned on having $K(t)$ active neurons at time $t$, for $4$ values of $\Delta \in \{ 20, 40, 80, 160 \} [\mathrm{ms}]$.  The trained RBM is the same as in Fig.~\ref{fig:genstats_SI}.}
    \label{fig:dynamics_SI} 
\end{figure*}
\begin{figure*}[h!]
    \centering
\begin{overpic}[width=0.9\textwidth]{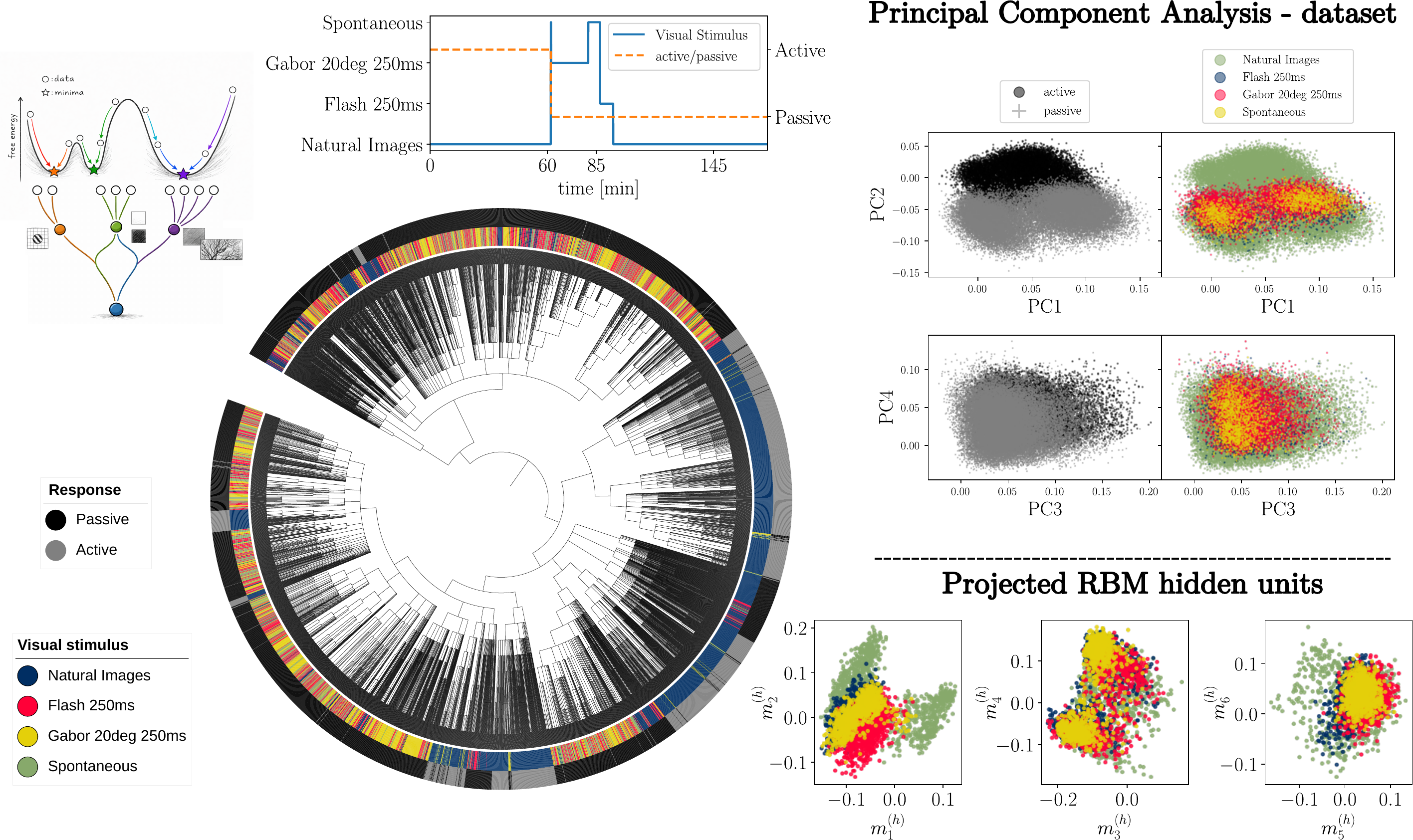}
\put(32,53){{\figpanel{a}}}
\put(63,53){{\figpanel{b}}}
\put(62,17){{\figpanel{c}}}
\put(10,35){{\figpanel{d}}}
\end{overpic}
    \caption{\textbf{Inferring activity/stimuli from the free-energy landscape.} \figpanel{a} Timeline of the experiment showing the presented visual stimuli and the mouse behavioral state during active recognition (licking task) and passive replay phases.  \figpanel{b} Neural activity projected onto the first four principal components, with data points coloured according to the label associated with each time point, illustrating that inferring the labels from this linear low-dimensional representation is a highly nontrivial task. Instead, in \figpanel{c} we project the hidden magnetizations  states associated to each data-point projected on the first two singular vectors of RBM's weight matrix where projections now split the different colourled datasets. In \figpanel{d} we construct a tree from the TAP fixed points at different stages of training, using 1000 samples from each label and following the procedure of Ref.~\cite{decelle2023unsupervised}. The resulting structure shows that TAP approximations of the free-energy minima naturally cluster neural activity patterns with the same label, and are even able to correctly classify some categories despite having no access to label information during training.  The trained RBM is the same as in Figs.~\ref{fig:genstats_SI},~\ref{fig:dynamics_SI}.
    }
    \label{fig:fixed-points_SI}
\end{figure*}

\begin{figure*}[h!]
    \centering
\begin{overpic}[width=\textwidth]{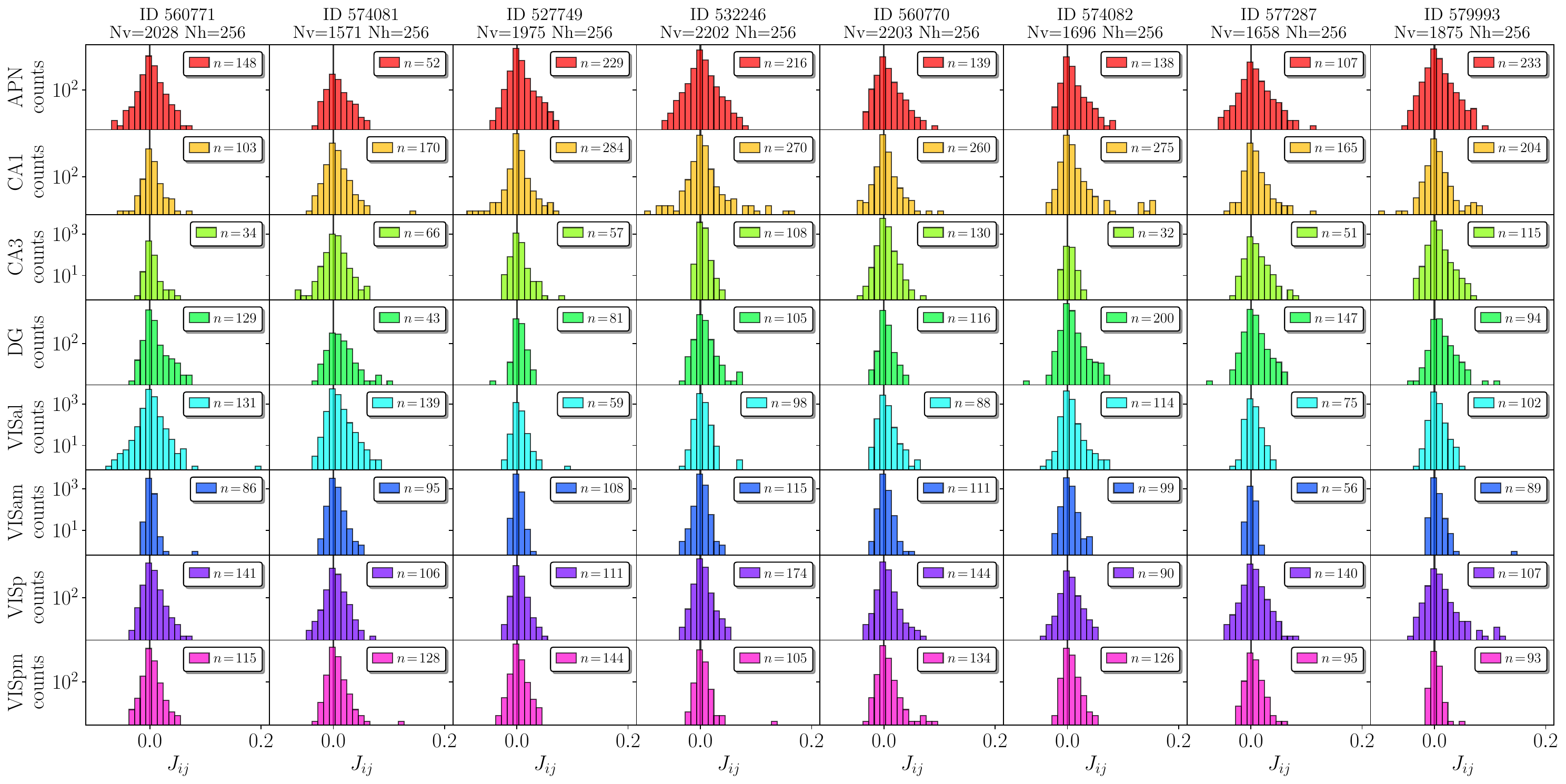}
\end{overpic}
    \caption{
    Histograms of intra-area two-body interactions $\{J_{ij} \mid  i,j \in A\}$. Each column refers to one mouse and each row to one of the $8$ regions with at least $30$ recorded neurons in all mice. The couplings are computed at the early-stopping time that maximizes the test log-likelihood. }
    \label{fig:histograms_Jba}
\end{figure*}

\subsection{Region-resolved distribution of two-body interactions}

In Fig.~\ref{fig:histograms_Jba} we report the histograms of the intra-area two-body
effective couplings $\{J^{(2)}_{ij}\mid i,j\in A\}$, resolved for each of the eight
brain regions containing at least $30$ recorded neurons in all mice (rows) and
for each of the eight experimental sessions (columns). The couplings are
extracted at the early-stopping time that maximizes the test log-likelihood.
Across all regions and animals the distributions are strongly peaked around
zero and display a pronounced positive tail (note the logarithmic vertical
scale), consistent with the sparse, heavy-tailed organization of effective
interactions already observed at the global level in the main text. This
confirms that only a small fraction of neuron pairs within a region sustain
strong direct couplings, while the bulk of the intra-area interactions are
weak.

A closer inspection reveals a clear heterogeneity across regions:
some areas exhibit visibly broader, more fat-tailed coupling distributions than
others, indicating that the strength and sparsity of the inferred interaction
network is region-dependent rather than uniform across the brain. Part of this
variability correlates with the number of recorded neurons per region (reported
as $n$ in each panel), so we cannot fully disentangle an intrinsic regional
property from a sampling effect, in line with the caveats discussed in the main
text. Importantly, for a fixed region the shape of the distribution is
qualitatively reproduced across different animals, supporting the conclusion
that the RBM captures consistent, animal-independent statistical features of the
underlying microscopic interactions.

\subsection{Degree distributions of the thresholded two-body network}
\begin{figure*}[h!]
    \centering
\begin{overpic}[width=\textwidth]{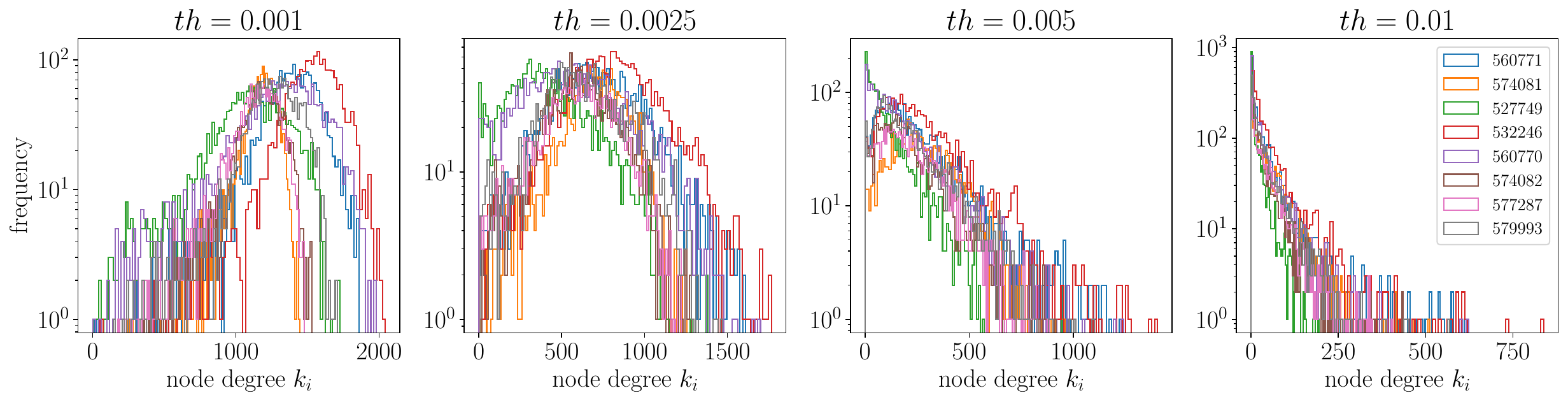}\end{overpic}
    \caption{Degree distributions of the two-body effective network obtained by thresholding the two-body interactions (in absolute values) to a given threshold, that is, the degree distributions of the adjacency matrix obtained by $\mid J_{ij} \mid> \mathrm{th}$. Each of the $4$ panels corresponds to a different threshold.
    }
    \label{fig:degree_distributions}
\end{figure*}

To further characterize the topology of the inferred interaction network, in
Fig.~\ref{fig:degree_distributions} we show the degree distributions of the two-body effective
network obtained by thresholding the couplings in absolute value, i.e. by
retaining only the edges satisfying $|J_{ij}| > \mathrm{th}$ and computing the
node degree $k_i$ of the resulting adjacency matrix. The four panels correspond
to increasing thresholds $\mathrm{th} \in \{0.001,\,0.0025,\,0.005,\,0.01\}$,
and the different colors denote the eight mice.

At the smallest threshold the network is still densely connected, with node
degrees reaching values comparable to the system size and a broad,
bell-shaped degree distribution. As the threshold increases, the network
progressively sparsifies: the typical degree drops by roughly an order of
magnitude and the distribution becomes markedly right-skewed, developing a
heavy tail in which a minority of neurons retain a large number of strong
connections while most nodes have low degree. This crossover from a dense,
weakly-interacting regime to a sparse, hub-dominated one is a direct
manifestation of the heavy-tailed coupling statistics of Fig.~\ref{fig:histograms_Jba}.
Remarkably, the degree distributions largely overlap across the eight animals
at every threshold, confirming that the coarse topological organization of the
inferred interaction network is a robust and reproducible property of the
RBM-based inference rather than an artifact of a particular session.

\end{document}